\documentclass[aps,preprint]{revtex4}
\usepackage{tabularx}
\usepackage{graphicx}
\usepackage{bm}
\usepackage{amsfonts}
\usepackage{amsmath}
\usepackage{exscale}
\usepackage{amsbsy}
\def\prb{Phys. Rev. B}
\def\prl{Phys. Rev. Lett.}

\def\be{\begin{equation}}
\def\ee{\end{equation}}
\def\ba{\begin{eqnarray}}
\def\ea{\end{eqnarray}}
\def\ess{\varepsilon}
\def\rr{{\bf r}}
\def\rrr{{\bf r}^\prime}

\begin{document}

\title{Theory of quantum metal to superconductor transitions in highly conducting systems}
\author{B. Spivak}
\affiliation{Department of Physics, University of Washington, Seattle WA 98195, USA}
\author{P. Oreto, S. A. Kivelson}
\affiliation{Department of Physics, Stanford University, Stanford CA  94305, USA}

\begin{abstract}
We derive the  theory of the quantum (zero temperature) superconductor to metal
transition in disordered materials when
 the resistance
of the normal metal near criticality is small compared to the quantum of resistivity. This can occur most readily in situations in which
 ``Anderson's
theorem'' does not apply.
We  explicitly study the transition in superconductor-metal composites,
 in an s-wave
superconducting film in the presence of a  magnetic field, and in a low
temperature disordered  d-wave
superconductor. Near the
point of the transition, the  distribution of the
superconducting order parameter is highly inhomogeneous.
To describe this situation we employ a procedure which is similar to that introduced by Mott for description of the temperature dependence of the variable range hopping conduction. As
the system approaches the point of the transition from the metal to the superconductor,
 the conductivity of the system diverges, and the
Wiedemann-Franz law is violated. In the case of d-wave (or other exotic)
superconductors we predict the existence of
(at least) two sequential transitions
 as a function of increasing disorder: a d-wave to s-wave, and then an s-wave to metal
transition.
\end{abstract}
\date{March 15, 2008}
\maketitle

\section{Introduction}

The  quantum (zero temperature) transition from a superconducting to
a non-superconducting ground state is the poster child of quantum
phase transitions.  This transition
is induced by
changing external parameters at zero temperature $T$.

In this article  we consider three representative problems in which a direct quantum phase transition
occurs from a superconducting
to a  metallic phase
in which $k_F l \gg 1$:
the case of a composite of a superconducting and a non-superconducting metal in which the effective interaction
between electrons changes sign as a function of position (Section II), the case when s-wave superconductivity is destroyed by
an external magnetic field (Section III), and the case when d-wave (or other exotic) superconductivity is
destroyed by quenched disorder (Section IV).  Here $k_{F}$ is the Fermi wave-vector, and $l$ is the electron mean free path.

In three dimensions (3D), there is no question concerning the existence of both a superconducting and a metallic phase.  In 2D,  the existence of a metallic phase is problematic.  However,
for $k_F l \gg 1$, single particle localization occurs on such large length scales that its effects are mostly unobservable.  Therefore, for most of this article, we will ignore the fundamental, but for our purposes  purely academic
  question of  whether or not a 2D interacting system of electrons can ever exhibit a metallic state in the asymptotic limit of zero temperature and infinite volume.

Before proceeding to discuss the  findings of the present study, it is worth commenting briefly on the broader context.  The central insight underlying the modern theory of critical phenomena is that, due to the divergent correlation length at criticality, the properties of the critical state are ``universal.''  An important extension of this is the idea that, in systems with quenched disorder, the variations in local environments are self-averaging, so a near-critical system can be treated   in terms of an effective, translationally invariant field theory \cite{HarrisLubensky,Khmelnitskii}. While this approach has had notable successes for the theory of classical phase transitions, it is  more problematic in the case of quantum critical phenomena.   This is most dramatically illustrated by the case\cite{dfisher,huse,young} of the quantum critical point in the random transverse-field Ising model, where the physics of ``rare events'' results in the existence of  a ``quantum Griffith phase'' in which, for a finite interval of parameters including the critical point, the susceptibility diverges as $T\to 0$.   For somewhat analogous reasons,
 a generic feature that characterizes the transitions in all three cases mentioned above  is that, at
criticality, the spatial distribution of the superconducting
 order parameter is highly inhomogeneous.
It  is concentrated  in
 ``superconducting puddles'' where, due to randomness, superconducting
order is locally anomalously favorable, and the distance between ``optimal'' puddles is parametrically large.
The transition occurs when  the Josephson coupling between optimal puddles (which falls with a power of the separation) times the exponentially large local superconducting susceptibility on a puddle is strong enough to stabilize a macroscopically phase coherent state.

Near enough to the quantum phase transition and at low enough temperatures, where the correlation length is large compared to the distance between superconducting puddles, there is still,
presumably, universal behavior described by appropriate critical exponents.  However, a consequence of the large distance between optimal puddles is that this universal quantum critical regime is parametrically narrow.  Conversely, there is an anomalously broad portion of the phase diagram in which the correlation length is comparable to or smaller than the distance between optimal puddles, but large compared to the size of an optimal puddle, where quantum fluctuations of the superconducting order-parameter dominate much of the physics.  This broad quantum but not quantum critical regime is one of the  characteristic signatures of the inhomogeneous nature of the critical state.

It is obvious that the presence of significant  superconducting
correlations in the ``metallic'' state near to the superconductor to metal critical point
makes it highly anomalous: Its zero temperature conductivity
diverges at the point of the transition, and can be much larger than the Drude value ``near'' criticality. In the same regime, the Hall conductivity decreases with
 respect to the Drude value and vanishes at the point of the transition.
 The Wiedemann-Franz law  in such metals is also clearly violated.

One remarkable implication of the present analysis is that, in the case when d-wave superconductivity
is destroyed by disorder, there are (at least) two quantum transitions: the first from a globally d-wave to a globally s-wave (although, possibly, still locally d-wave) state, and the second to the ``normal'' metal.
 Another outcome of this picture is peculiar temperature dependencies
of the physical parameters of the near critical superconductor. Some of our findings are summarized in the schematic phase diagrams shown in  Figs. 1-3.

To conclude this introduction, we would like to discuss the relation between our paper and those
\cite{Fisher,Girvin,goldman} in which  it has been proposed that the quantum transition, especially in 2D, takes place between the superconducting and an insulating state.
Several  lines of reasoning led to the inference that near the $T=0$ quantum critical point $k_{F}l\approx 1$ , in which case localization (which we neglect) would necessarily be a serious issue:

1) Where  Anderson's theorem applies at the level of mean-field theory, such as in the case in which s-wave superconductivity
 is destroyed by increasing disorder, the localization length must be comparable to or shorter than the coherence length in order for the disorder  to have   any substantial effect on $T_c$, at all \cite{kotliar}.

  2) It has been shown that  in a system of
superconducting grains linked by resistively shunted Josephson
junctions, quantum fluctuations of the order parameter \cite{Luther,Zimanyi, Kivelson,Fisher} are strongly suppressed so long as $G^{(eff)}\gg 1$, independently of the strength of the Josephson coupling between puddles! Here $G^{(eff)}$ is a dimensionless  shunt conductance measured in units $e^{2}/h$.
An apparent implication of this result is that,  so long as a small portion of a highly conducting system is superconducting, at low enough temperature the system will achieve global phase coherence so long
as the dimensionless effective conductance, $G^{(eff)} $, is large compared to 1.

3) It was found in  several
theoretical studies \cite{Fisher,Girvin,goldman}  of 2D bosonic models of the transition  that there is a universal value of $G^{(eff)}=G_c\sim {\cal O}(1)$ at the point of superconductor-metal transition.  (These models assume the absence of gapless quasiparticle excitations, and therefore ignore dissipation of the sort that is represented by the shunt resistance in the previously discussed models.)

4) A large portion of the experimental realizations of such transitions involve two dimensional (2D) systems, such as films.  Here, the transitions are often referred to as a ``superconductor to insulator transition'' \cite{Fisher} on the basis of the widely held theoretical belief that metallic phases are forbidden  in 2D due to single particle localization \cite{Gangogfour,KhmLarkGor} -- any non-superconducting phase is thus expected be insulating at zero-temperature.

 Concerning point 1),
 in the present paper, for the most part, we consider problems in which Anderson's theorem does not apply, either due to the symmetry of the order parameter ({\it e.g.} d-wave) or the breaking of time reversal symmetry ({\it e.g.} by an applied magnetic field).
Concerning point 3),
we consider cases in which gapless quasiparticles are present near criticality, so the applicability of a bosonic model is questionable.
Concerning point 4), as mentioned above, 2D localization is negligible, and hence a non-superconducting phase is ``effectively metallic,'' whenever the parameter $k_{F}l$ is sufficiently large,
since in this limit, the localization length is exponentially large, $\xi_{loc}\sim l\exp(\frac{\pi}{2}G_{2D})$.
Here $G_{2D}\propto k^{2}_{F}l d \gg 1$ is  the dimensionless conductance per square
measured in units of $e^{2}/h $ of a 2D film of thickness $d$.
(For a review, see  \cite{LeeRamakrishnan}.)  Finally, concerning point 2),
a large portion of the discussion in the present paper follows from the same considerations.   The differences between the present results and those of the earlier studies spring from the fact that the effective model we develop from microscopics differs in a  subtle manner from  the phenomenological Ohmic heat bath considered in those earlier studies.  This difference permits a transition to a phase incoherent state even under conditions in which $G^{(eff)} \gg 1$;  however, a residual consequence of the same physics discovered in those earlier studies is that this transition occurs when the superconducting puddles are extremely dilute and so are weakly Josephson coupled to one another.  This is one of the central  results of the present study.

\subsection{Effective action for the quantum superconductor-metal transition}

In the present section, we develop the general features of the effective actions that govern the quantum fluctuations of the order parameter near criticality.  Formally, such effective actions are obtained by integrating out  the fermionic degrees of freedom, and all high energy collective modes, leaving us with a set of degrees of freedom, $\Delta_i = |\Delta_i|\exp[i\phi_i]$, identified as  the phase and modulus of the order parameter on puddle $i$. In detail, the various terms are sensitive to the specific physical circumstances, but the overall structure of the effective action is the same in all cases studied in the present paper.

In the case of small puddles embedded in a  normal metal,  where the value of the order parameter is small, the Andreev reflection of quasiparticles from the metal-superconductor boundary
is ineffective and can be taken into account in perturbation theory.
Then, the imaginary time effective action that governs the superconducting
 fluctuations near criticality is of the form
\ba
S= \sum_j & \Bigg\{  &\alpha_j \int d \tau \left[ -\frac {
(\gamma-\gamma_{jc})}
 {2}|\Delta_j|^2 + \frac {1} 4 \frac{ |\Delta_j|^4}
 {\Delta_0^2} \right]  \nonumber \\
&&+\frac {\beta_j} 2  \int d \tau d \tau^\prime
\frac{|\Delta_j(\tau)-\Delta_j(\tau^\prime)|^{2}}
{(\tau-\tau^\prime)^{2}}\Bigg\} +\int d \tau H_{J}[\{\Delta\}] +\ldots
\nonumber \\
&& H_{J}[\{\Delta\}]= -(1/2)\sum_{i\neq j}  [J_{ij} \Delta_i^\star\Delta_j +
{\rm c.c.}]
 \label{TDGL}
\ea
where $\tau$ is imaginary time,  $j$ labels the randomly distributed superconducting
puddles, $\gamma$ is the parameter that tunes the phase transition
 ({\it e.g.} the magnetic field in units of the critical magnetic field), $\gamma_{jc}$ is the critical value of
 $\gamma$ in the $j_{th}$ puddle  ,  $\Delta_0$ is the
magnitude of the order parameter deep in the superconducting state ,
$\alpha_j$ and $\beta_j$  depend on the local structure of the
 superconducting puddle (as discussed below), and $J_{ij}$ is the
 Josephson coupling between nearby puddles.  The $\ldots$ symbol represents high order
terms in $\Delta$ that are negligible at the phase transition, including non-local quartic terms involving the order parameter on more than one puddle. (Some representative aspects of the derivation of Eq.~\ref{TDGL} are sketched in the Appendix.)

The first and the third terms in Eq.~\ref{TDGL} reflect the dynamics of the BCS Cooper instability
\cite{LarkinGalitskii,Herbut,HruskaZyuzinSpivak}, and hence
\begin{equation}
\alpha_{i}\sim \nu V_{i}, \,\,\,\, \beta_{i}\sim
\nu V_i/{\Delta_{0}}
\label{coefficients}
\end{equation}
where $V_i$ is the volume of the $i^{th}$ puddle and $\nu$ is the density of states in the surrounding metal.  (In two dimensional cases, naturally, $V_i$ is the area of the puddle and $\nu = \nu_{3D} d$ is the
two dimensional density of states.)
To illustrate this, set $J_{ij}=0$ and consider the dynamics of small amplitude fluctuations of the order parameter on an isolated puddle
 that is on the verge of becoming superconducting ($1\gg (\gamma_{ic}-\gamma)>0$) as described by the linearized version of Eq. \ref{TDGL}:
\begin{equation}
\chi_i(\omega) \equiv \int dt e^{i\omega t} \langle \Delta^{*}_{i}(0)\Delta_{i}
(t)\rangle= \frac{1}{2\pi (\beta_j|\omega|+1/\tau_{i})}; \,\,\,\,\,\,\,\,\
\frac 1 {\tau_{i}}=\frac{\alpha_i(\gamma-\gamma_{ic})} {2\pi}.
\label{CupDecay}
\end{equation}
Comparing this with the usual calculation of gaussian fluctuations in the neighborhood of the superconducting transition leads to the expressions in Eq. \ref{coefficients}, {\it i.e.} Eq. \ref{CupDecay} simply describes the dynamical fluctuations which lead to the Cooper instability as $\gamma\rightarrow \gamma_{ic}$.
Here  brackets $\langle\ \  \rangle$  signifies the quantum expectation value.

In the opposite limit, when the puddles  are large with big values of the order parameter,
one can neglect quantum fluctuations of the modulus of the order parameter
and write the effective action in terms of fluctuations of the phase only
\ba
&&S=
  \sum_{j} G^{(eff)}_{i}  \int d \tau d \tau^\prime {\left
|e^{i\phi_{j}(\tau)}- e^{i\phi_{j}(\tau^{\prime})}\right|^2\over
(\tau-\tau^\prime)^2} + \int d\tau H_J[\{\phi\}] \nonumber \\
&&H_J[\{\phi\}]=-(1/2)\sum_{i\neq j} \tilde{J}_{ij} \cos(\phi_{i}-\phi_{j})
 \label{TDGL1}
\ea
The form of the effective action in Eq.~\ref{TDGL1} is familiar from
many earlier studies of the quantum dynamics of a system of
superconducting grains linked by resistively shunted Josephson
junctions (See, for example, Refs.
\onlinecite{Chakravarty,Schmid,ambagoakar,Luther, Fisher,Kivelson,Girvin}.)  In particular,
the dynamical term proportional to
$G^{(eff)}_{i}$ in Eq. \ref{TDGL1} has  the familiar Caldeira-Leggett \cite{leggett} form
and  describes the quantum dynamics of the order parameter of an
isolated puddle. In this case the origin of the dynamical term
in Eq. \ref{TDGL1},
is entirely different from that in Eq. \ref{TDGL}:
it reflects the interaction of
 the phase fluctuations of the superconducting order parameter
with quantum fluctuations of the electromagnetic field. In this case  $G^{(eff)}_{i}$ is the dimensionless effective conductance defined by injecting current into the i-th superconducting puddle embedded in the metallic host (ignoring the effect of other puddles), and measuring the voltage drop at infinity.
There is a further subtlety in 2D, as discussed in Refs. \cite{feigelman,feigelmanSkvortsov}, associated with the fact that, $G^{(eff)}_{i}$, as so defined, vanishes logarithmically with the size of the system.  To handle this problem properly, one needs to consider corrections to the dynamical term in the effective action, Eq. \ref{TDGL1}.  When this is done, the result is  equivalent to identifying $G^{(eff)}_{i} \sim\sqrt{G_{2D}}$.

We will show below that in different situations either electromagnetic fluctuations or the Cooper  instability can make the dominant contribution to the quantum dynamics of the order parameter .

\subsection{Josephson couplings between puddles}

We will see that near criticality the typical inter-puddle distance is larger than their size.
The other generically important aspect of the problem is the dependence of the Josephson
couplings between puddles on their separation, ${\bf r}_i - {\bf r}_j$, which is
long-ranged (power-law)  in the limit  that the temperature, $T\to 0$.  Specifically, the coupling   between small puddles in Eq. \ref{TDGL} is, up to logarithmic corrections which we will discuss later,  of the form
\begin{equation}
J_{ij}\equiv J({\bf r}_{i},{\bf r}_{j})\propto  C_{ij}\frac{ \nu V_{i}V_{j}} {|{\bf r}_{i}-{\bf
r}_{j}|^{D}}\ \exp\left[-\frac{|{\bf r}_{i}-{\bf r}_{j}|}{L_{T}}\right],
\label{josephson}
\end{equation}
where $\nu$ is the density of states in the normal metal,
 $L_{T}=\sqrt{D_{tr}/T}$ is the coherence length of normal metal  which diverges as $T\to 0$, $D_{tr}$
is the electron diffusion coefficient, and  $C_{ij}({\bf r}_{i},{\bf r}_{j})$ is, generally speaking, a complicated (random) dimensionless function of the coordinates.  In the cases considered in Secs. II and IV,  the value of $C$ is determined by the average properties of the ``normal'' metal between puddles.
 In particular, in  Sec. II, $C_{ij}$ is mostly positive, and so can be approximated by its average value, $\overline{C_{ij} }$, while for the d-wave superconductor treated in Sec. IV, $C_{ij}$ has a random sign, but at long distances, this sign is entirely determined by the character of the puddles, $i$ and $j$, and is independent of the distance between puddles.  By contrast,
 in the cases with a magnetic field considered in Sec. III ,
 $C_{ij}$ is determined by the random quantum interference between different paths through the normal metal.  As a consequence,  $C_{ij}$ has a random phase.

In the limit of large puddles, the functional dependence of the Josephson coupling in Eq.~\ref{TDGL1},
\begin{equation}
 \tilde{J}_{ij}\propto \tilde{C}_{ij}\frac{D_{tr}}{R^{2}}\frac{R^{D}}{|{\bf r}_{i}-{\bf r}_{j}|^{D}}\exp\left[-\frac{|{\bf r}_{i}-{\bf r}_{j}|}{L_{T}}\right]
\label{J_ijpm1}
\end{equation}
on distance is the same as in Eq. \ref{josephson}. Here the random function $\tilde{C}_{ij}$ has properties similar to  $C_{ij}$, and
 $R$ is the typical size of the puddles.

\subsection{Susceptibility of an individual puddle}

The susceptibility of an individual puddle can be expressed in terms of the correlation function
$\chi_i(\omega)$  (Eq. \ref{CupDecay}) as $\chi_i\equiv \chi_i(\omega=0)$.
 Its value depends on the puddle size.

 Let us start with the case when the puddle is large,
 so the dynamics of the order parameter is determined by the
 effective action given by Eq.~\ref{TDGL1} with $J_{ij}=0$.
The implications of this effective action are
best appreciated by interpreting imaginary time as a fictive spatial
dimension, making
the single puddle problem equivalent to
the classical inverse-square XY model \cite{kosterlitz,zwerger} at
an effective temperature $T^{eff} = 1/G^{(eff)}_{i}$. The
long-time correlation functions of the inverse X-Y model have been
calculated in various ways, and are well understood.  The characteristic
decay time depends exponentially on $1/T^{eff}$, and the dynamic
correlation function has a power-law fall-off \cite{CandL},
\begin{equation}
\langle \Delta^{\star}_{i}(0) \Delta_{i}(\tau)
\rangle
  \sim |\Delta_j|^2
\left\{
\begin{array}{ccc}
\left[
{\tau_{j}}/{\tau}\right]^{x_j}  &  {\rm for} &  \tau \ll \tau_j   \\
\left[
{\tau_{j}}/{\tau}\right]^{2}  &  {\rm for} &   \tau \gg \tau_j
\end{array}
\right.
  \label{decay}
\end{equation}
where $x_i$ is a non-universal exponent
 $x_i= T^{eff}/(2\pi)$, the relaxation time,
 \begin{equation}
 \tau_i\sim \exp[
 ZG^{(eff)}]
 \label{decay2}
 \end{equation}
 and $Z=2\pi^2$.
The susceptibility is thus
\begin{equation}
\chi_i \sim \Delta_0 \exp[ ZG^{(eff)}_{i} ].
\label{3D}
\end{equation}
However,  in 2D, due to the dimension specific subtlety \cite{feigelman,feigelmanSkvortsov} discussed in Subsection IA, we must identify $G^{(eff)}_{i} \sim \sqrt{G_{2D}}$, so that
\begin{equation}
\chi_{i}\sim \Delta_{0} \ \exp\left[ \ Z^\prime \sqrt{G_{2D}} \  \right]
\label{FeigLarkSusc}
 \end{equation}
 where $Z^\prime$ is another number of order 1.

Let us now turn to the case when the modulus of the order parameter on a puddle is small and its dynamics are determined
by Eq.~\ref{TDGL}.
 There is a complicated, and for our purposes not terribly important crossover that occurs for puddles which are right on the verge of a mean-field transition, $|\gamma-\gamma_{ic}| \ll 1$.
Ignoring such puddles, there are two distinct asymptotic behaviors that are readily deduced:
\begin{equation}
\chi_i \sim
\left\{
\begin{array}{ccc}
[{\alpha_i(\gamma_{ic}-\gamma)}]^{-1}  &   {\rm for}  &  (\gamma_{ic}- \gamma) > 0, \\
| \Delta_i|\ \exp[\ Z''\beta_i|\Delta_i|^2\ ]
  & {\rm for}  &  (\gamma_{ic}- \gamma) < 0, \
\end{array}
\right.
\label{suscept}
\end{equation}
where,  in this expression, $|\Delta_i|=\Delta_0\sqrt{\gamma-\gamma_{ic}}$ is the mean-field amplitude of the superconducting order on puddle $i$ and $Z''$ is a
renormalized  relative of the same factor $Z$ defined in Eq. \ref{3D}, above.  We now sketch the derivation of this result.

The result for a puddle that does not support a mean-field solution, {\it i.e.} for $\gamma_{ic} > \gamma$,
 is easily obtained by evaluating  Eq. \ref{CupDecay} at $\omega=0$.
In the opposite limit, $(\gamma-\gamma_{ic})> 0$,
there is a well developed mean field value of the order parameter on the puddle
 \begin{equation}
 \Delta_{j}=\Delta_{0} \sqrt{
 {\gamma -\gamma_{jc}}
 }\exp[i\phi_j],
 \end{equation}
  which has an arbitrary phase $\phi_j$.  To the extent that modulous fluctuations can be ignored, this problem is precisely equivalent to the large puddle problem, with the role of $G^{(eff)}_{i}$ played by $\beta_j\Delta_j^2$ on each grain.  Although large amplitude modulus fluctuations are relatively costly, because the resulting expression for the susceptibility depends exponentially on $|\Delta_j|^2$,  they cannot  be neglected.  However,
 since the modulus of the order parameter appears exponentially in the
expression for $\chi_i$, it is clear that the neglect of modulus fluctuations is not reasonable.   However, they do not alter the asymptotic qualitatively, but rather  result in a renormalization (reduction) of the  factor $Z''$ in Eq. \ref{suscept}.

The most important feature of Eq. \ref{suscept} is that the susceptibility increases
exponentially as a function of $\gamma-\gamma_{ic}$ and of the volume, $V_i$, of the puddle, $\chi \sim \exp[ Z''\nu \Delta_0 V_i(\gamma-\gamma_{ic}) ]$.

 \subsection{Determination of the quantum critical
point}

We now outline the procedure for determination of the location of the quantum critical point under these circumstances.

Quantum fluctuations necessarily destroy the superconducting order in an isolated
puddle. Thus, although the   superconducting susceptibility of an  individual puddle, $\chi_{i}$, can, under some circumstances, be large, the transition to the globally phase coherent superconducting state is ultimately triggered by
the Josephson coupling between puddles. Let us introduce a dimensionless coupling between two puddles, $i$ and $j$,
\begin{equation}
 X_{i,j}
\equiv \chi_{i} J_{i,j}\chi_{j} J_{j,i} . \label{phasetrans}
\end{equation}
Two puddles fluctuate essentially independently of each other if $|X_{i,j}| \ll 1$, and they are phase locked to each other
if $|X_{i,j}| \gg 1$.  The transition to a globally phase coherent state occurs as a function of $\gamma$ at the critical value, $\gamma=\gamma_c$, at which an infinite cluster of puddles is coupled together by links with $X_{i,j}\sim 1$.
For an ordered array of puddles, the quantum superconductor-metal transition was discussed in this light
in \cite{feigelman,HruskaZyuzinSpivak,feigelmanSkvortsov}.

In disordered systems, the nature  of the phase transitions described by the effective action in Eq. \ref{TDGL} depends on the distribution of the parameters, $\gamma_{ic}$, $\beta_j$, $G^{(eff)}_{i}$  and $J_{ij}$, and these in turn are somewhat different in the various cases we treat below.

However, what is common to the cases we will analyze is that,
according to Eq.\ref{suscept} the susceptibilities of the puddles depend
exponentially on the parameters of the action Eq.\ref{TDGL}.
Thus in a
generic situation in the neighborhood of the transition, the distribution of $\chi_{i}$  is extremely broad, and at criticality, rare puddles with exponentially large
susceptibilities play a special role. In this case, the critical point can
be identified by finding
 the set of
``optimal puddles" which lie on the critical links of ``the
percolating cluster". This will be done  in a way analogous to
 Mott's approach to the theory of the variable range hopping conductivity
 (See for example \cite{ShklovskiiEfros}).

  Specifically, the optimal puddles are those in which
 $\gamma_{ic}$ lies in an interval,
 $\gamma_{opt}-\Delta\gamma_{opt} < \gamma_{ic} < \gamma_{opt}+\Delta\gamma_{opt} $. Here both the optimal value, $\gamma_{opt}$, and the width of the interval, $\Delta\gamma_{opt}$, are determined
by maximizing the quantity $X_{opt}=\chi_{opt}^{2}J_{opt}^{2}$ with respect to these parameters, where $\chi_{opt}$ is the susceptibility of a puddle with $\gamma_{ic}=\gamma_{opt}$, and $J_{opt}$ is the typical value of the Josephson coupling between two nearest-neighbor optimal puddles. Finally we find the critical value of $\gamma=\gamma_{c}$ from the requirement that, after maximizing, $max \{X_{opt}\}\approx 1$ . In the following sections we consider several examples of this program.

\section{A random mixture of metal and superconductor}

As a first case, we
consider a random set of s-wave superconducting grains  is  embedded in a
normal metal host with no magnetic field or magnetic impurities.
We identify the superconducting grains as regions in which the effective interaction between two electrons in the Cooper channel is attractive ($\lambda_{S}>0$), while in the normal metal the interaction is repulsive ($\lambda_{N}<0$).
This system
exhibits a metal-superconductor
transition when the appropriate average value of $\lambda({\bf r})$ changes sign, although
the parameter $k_{F}l$ can still be arbitrarily large
\cite{feigelman,HruskaZyuzinSpivak,feigelmanSkvortsov}. Some aspects of various closely related problems have been previously been analyzed using a variety of approaches\cite{feigelman,HruskaZyuzinSpivak,feigelmanSkvortsov,oreg,voigta,sachdev,motrunich}.  However, we are able to obtain a much complete picture than has been obtained previously.  Moreover, this problem serves as a useful warmup as it provides a simple explicit example of how the character of the optimal puddles and the nature of the quantum phase transition are determined from the present quantum percolation analysis.

 To be concrete, we will assume that the
diameters of the grains, $R_{i}$, are random quantities characterized
by the Gaussian distribution
\begin{equation}
P(R_{i})=\frac{N}{\sqrt{2\pi}\sigma_{R}\bar R}\exp\left[-\frac{(R_{i}-\bar{R})^{2}}
{2\sigma^{2}_{R}\bar R^2}\right]
\end{equation}
where the average radius is $\bar{R}$, the dimensionless variance
$\sigma_{R}\ll 1$, and the total concentration of grains is
$N$. In the notation of the previous section, we can
identify $\gamma_{i} - \gamma
=  R_{i}/R_{c}-1$,
where
$R_{c}\sim \xi_{0}$ is the critical radius for the existence of a mean-field solution and
$\xi_{0}$ is zero temperature coherence length in the superconductor.

Expressions for the susceptibilities of individual grains, in various limits,
are given in  Eqs. \ref{3D}, \ref{FeigLarkSusc}, and \ref{suscept}.

The value of the Joshepson coupling between two
superconducting grains embedded in a normal metal
depends on whether the Andreev reflection on the N-S boundary is effective or not.
When the puddles are larger than the coherence length determined by the
magnitude of the order parameter in the puddle, one finds
\begin{equation}
\tilde{J}_{ij}\sim {G}_{eff}\frac{D_{tr}}{\bar{R}^{2}}\frac{\bar{V}}{|{\bf r}_i-{\bf r}_j|^{D}[1+2\lambda_{N}|\ln^{2}(|{\bf r}_{i}-{\bf r}_{j}|/\bar{R})|]}\exp\left[-\frac{|{\bf r}_i-{\bf r}_j|}{L_{T}}\right]
\label{J_ijpm1}
\end{equation}
 by solving the Usadel equation. (See for
example \cite{likharev}). Here  $\bar V$ is the volume of a grain,  $\bar V \sim \bar R^D$.
In the opposite limit, when the value of the order parameter on the puddle is small and Andreev reflection is ineffective, the coupling can be computed from perturbation theory (See for example \cite{HruskaZyuzinSpivak}:
\begin{equation}
 J_{ij}\sim
 \frac{\nu
 \bar V^2}
 {|{\bf r}_{i}-{\bf r}_{j}|^{D}\left[1+2|\lambda_{N}
|\ln^{2}(|{\bf r}_{i}-{\bf r}_{j}|/\bar{R})
\right]} \exp\left[-\frac{|{\bf
r}_{i}-{\bf r}_{j}|}{L_{T}}\right] ,
\label{J_ijpm}
 \end{equation}.

Due to the s-wave character of the superconducting order and the fact that the system is time-reversal invariant, and so long as certain effects of strong correlations in the metal \cite{SpivakKivelsonNegative}
can be ignored, $J_{ij}$ is always real and positive. Moreover, if
 $k_{F}l\gg 1$, the mesoscopic
fluctuations of the magnitude of
 $J_{ij}$ are small compared to the average, and
 can thus be neglected.

Our goal is to determine the critical concentration of grains $N_{c}$ at which the superconductor-metal transition occurs.
We can identify
various regimes depending on values of the parameters $\sigma_{R}$, $(\bar{R}-R_{c})/R_{c}$,
 $G_{eff}$, and the Ginzburg parameter $g\equiv \nu R_c^D\Delta_{0}$.
($g$ is roughly the number of electrons within energy window $\Delta_0$ on an individual superconducting grain.
 Notice that, in the cases of interest here, where $R_c \sim \xi_0$, it is always the case that $g \gg 1$.)  We analyze some representative cases:

{\bf A)  Monodispersed superconducting grains:} Let us start with the case where  $\sigma_{R}\to 0$,
so all puddles are essentially the same size $R_{i}\equiv \bar{R}$.
At $T=0$, it is simple to see that:

 {\noindent \bf i)} If $(R_{c}-\bar{R}) < 0$,
  individual puddles are not superconducting, even at mean-field level.  In this case, which corresponds to the  uniform ``Cooper limit,''  the superconducting transition is, to first approximation, mean-field like and it occurs when
  the average interaction strength,
  \begin{equation}
  \bar \lambda \equiv N\bar V \lambda_N - (1-N\bar V)|\lambda_N|
  \label{barlambda}
  \end{equation}
  changes sign (from attractive to repulsive).  Thus, the critical concentration is
 $N_c \bar V \sim |\lambda_N|/[\lambda_S + |\lambda_N|].$  This estimate neglects the spatial variations in the local concentration of superconducting grains;  even when $N$ is, on average, smaller than this mean-field critical value, there occur regions in which the concentration of grains exceeds this critical value.  These regions act as the superconducting puddles of a new level of analysis, which gives results similar to those discussed below in the case of larger $\sigma_R$.

 {\noindent \bf ii)}
  If $1 \gg (\bar{R}-R_{c})/\bar{R} > 1/g$, there is a well-developed mean-field order, $\Delta=\sqrt{(\bar R-R_c)/R_c}\Delta_0$ on each grain, but the order parameter has a magnitude small compared to $\Delta_0$.  In this case, Eq. \ref{TDGL}
 governs the dynamics, and we can make use of the expressions Eqs. \ref{suscept} and \ref{J_ijpm} for $\chi$ and $J_{ij}$, respectively.  Since the grains are typically a distance of order $N^{-1/D}$ apart, the dimensionless coupling between two neighboring grains is
 \begin{equation}
 X \sim \left\{g\bar R^DN\sqrt{(\bar R-R_c)/R_c}\exp[Z^{\prime\prime}g (\bar R - R_c)/R_c] \right\}^2.
\end{equation}
This coupling is larger than 1 when $N$ exceeds the critical density,
 \begin{equation}
 N_c \sim \frac{1}{g\bar R^{D}}
\left[\frac{\bar{R}-R_{c}}{R_{c}}\right]^{1/2}
\exp \left[-Z'' \ g\ \left(\frac{\bar{R}-R_{c}} {R_{c}}\right)\right] \,\,\,\,\,\,\, {\rm at}\,\,\,\,\,\,\, T=0 .
\end{equation}

{\noindent \bf iii)} If $ (R_{c}-\bar{R})/\bar{R}  \geq 1$,  the grains act, more or less, like pieces of bulk superconductor.  Here, the dynamics of the quantum fluctuations are governed by electric field fluctuations
as in Eq. \ref{TDGL1}, and consequently the same analysis leads to
\begin{equation}
N_{c}
\sim \left\{
 \begin{array}{ll}
\bar R^{-2}\exp(-Z^\prime\sqrt{G_{2D}}) & \mbox{{\rm in} \ $D=2$ \ {\rm at} \ $ T=0$}
\\
\bar R^{-3}\exp(-ZG^{eff})  & \mbox{{\rm in} \ $D=3$ \ {\rm at} \ $ T=0$}.
\end{array} \right. \ \label{N_cord}
 \end{equation}

Let us turn now to the temperature dependence of $N_{c}(T)$.   Conventional arguments  suggest that at low enough temperatures, and arbitrarily close to the zero temperature critical point,  there is a universal quantum critical regime\cite{sachdevbook,comment}, where,  for example,  $(N_c(T)-N_{c})\sim N_c T^{x}$, where $x$ signifies an appropriate universal critical exponent.  This regime, if it exists, applies only so long as $L_T \gg N_c^{-1/D}$, and so is exponentially narrow.  Beyond the quantum critical regime, there is a broad range of temperatures in which
$N_c^{-1/D} \gtrsim L_{T}\gg \bar{R}$, where the fluctuations are highly quantum mechanical in the sense that one can still neglect the $T$-dependencies of $\chi(T)$.
In this case there are two sources of the $T$-dependence
of $N_{c}(T)$ : a) the classical fluctuations which destroy the coherence between puddles when
  $ J_{ij} |\Delta_{i}||\Delta_{j}|\approx T$, and b) the $T$-dependencies of $J_{ij}(T)$, given by Eq. \ref{J_ijpm}.
The relative importance of these two effects depends on the dimensionality of space $D$, and the value of the parameter $g$.
The second mechanism dominates the $T$-dependence of $N_{c}(T)$ in the 2D case at arbitrary $T$,
and in the 3D case in the wide interval of temperatures where $L_{T}/\bar{R}>g$.
Then
   the criterion $X_{opt}\sim 1$ corresponds to
  a typical distance between puddles of order $L_{T}$, and hence
\begin{equation}
N_{c}(T)\sim
\frac{1}{(L_{T})^{D}},  \,\,\,\,\,\,\,\ \bar R \ll L_{T} <  N_{c}(0)^{-1/D},
\label{N(T)}
 \end{equation}
 (In this article we will ignore relatively small temperature interval $L_{T}>\bar{R}$ where in 3D the temperature
 dependence $N_{C}(T)$ is determined by the first mechanism.)
 Notice, subtleties, such as whether the superconducting state has long-range order (the 3D case) or only power-law order (the 2D case), do not affect the validity of this estimate.

{\bf B) Optimal radius grains:} If the variance $\sigma_{R}$ is not too small, the susceptibilities of individual grains $\chi_{i}$ have
an exponentially broad distribution. As a result,  at $T=0$  the transition point occurs when  $\bar{R} <R_{c}$ and is determined by relatively rare ``optimal'' puddles
with anomalously large values of $(R_{i}-R_{c})/R_{c}$, and consequently with exponentially large
susceptibilities.
However, as we shall see,  if $\bar R$ is too much smaller than $ R_c$,  the optimal grains become so rare that, yet again, a new regime occurs  in which the optimal puddle is formed in regions with  an anomalously large concentration of sub-critical grains.

Let us
focus on those grains with radii, $R_{i}$, within $\Delta R_{opt}$ 
of a still to be determined optimal radius, $R_{opt}>\bar{R}$.
 (It can be shown that the relevant range is $\Delta R_{opt}\sim \sigma_{R}\bar{R}$).
We will ignore puddles which do not belong to this optimal set
since puddles with much larger $R_j$ are extraordinarily rare, and those with much smaller $R_j$ have much smaller susceptibilities.
Under the assumption (to which we will return  below) that the optimal puddles are still small enough that Eq. \ref{TDGL} applies, the concentration of the optimal puddles is
\begin{equation}
N_{opt}\sim N\sigma_{R}\exp[-(R_{opt}-\bar{R})^{2}/2\sigma^{2}_{R}\bar{R}^{2}]
\end{equation}
so,
according to Eqs. \ref{suscept},\ref{J_ijpm} and \ref{phasetrans}
\begin{equation}
X_{opt}\sim \left[N_{opt}g \  \right]^2
\exp\left[Z''g\frac{(R_{opt}-R_{c})}{\xi_0}\right] \label{Xopt}.
\end{equation}
Maximizing Eq. \ref{Xopt} with respect to $R_{opt}$ gives
\begin{equation}
R^{(max)}_{opt}=\bar{R}+Z''g\bar R^2\sigma^{2}_{R}/\xi_0
\label{optimalR}
\end{equation}
 and
 $max \{N_{opt}\} \sim N\exp\left[-(Z''g\bar R\sigma_{R})^{2}/2\xi_0^2\right]\ll N$ ({\it i.e.} most puddles are smaller than $R_{opt}$ and hence play no direct role in the transition).  Finally,  using the criteria
 $X_{opt}\sim 1$, we find
\begin{equation}
N_{c}\sim \frac{1}{\bar{R}^{D-1}\sigma_R}
\exp\left[\frac{Z''g}{\xi_0}\left(R_{c}-\bar{R}- \frac{Z''g\bar R^2\sigma_{R}} {2\xi_0}\right)\right]
\,\,\,\ {\rm at} \,\,\ T=0.
\label{CritConzPMDis}
\end{equation}
We note that  the critical concentration in Eq.~\ref{CritConzPMDis} can be extremely small
as a consequence of the fact that rare larger than average
puddles
contribute significantly to the
global phase coherence.

Let us now discuss the limits of applicability of Eq. \ref{CritConzPMDis}.
It is manifestly necessary that $\sigma_{R}$
be
small enough that $N_c \bar R^D \ll 1$, {\it i.e.} that
\begin{equation}
  (R_c-\bar R) > Z''g\bar R^2\sigma_R^2/2\xi_0.
\end{equation}
Note that that the same criterion leads to the inequality $\Delta
R_{opt}\sim \bar R\sigma_{R}\ll (R_{opt}-\bar{R})$, which justifies the saddle point approximation.

The temperature dependence of $N_c$ can be obtained in similar ways as in  {\bf A}, above.  When $L_T \gg N_c^{1/D}$, the behavior is presumably governed by universal properties of the quantum critical point.
However, for $N_c^{1/D} > L_T \gg \bar R$, the temperature dependence of $N_c$ again derives from the temperature dependence of $J_{ij}$, {\it i.e.} that the Josephson coupling falls rapidly to zero at distances large compared to $L_T$.
Here,
the character of the optimal puddles is still determined by Eqs. \ref{Xopt} and \ref{optimalR}, but the critical concentration is determined by the condition that the distance between optimal puddles is typically of order $L_T$,
{\it i.e.}

\begin{equation}
N_c(T) \sim L_T^{-D} \exp\left[\left(Z''n_c\bar R\sigma_{R}\right)^2/2\xi_0^2\ \right].
\end{equation}
The resulting phase diagram is shown schematically in Fig. 1.

{\bf C) Large puddles:}  The expressions in {\bf B} were derived under the assumption that the optimal grains are sufficiently small that the mean-field order parameter has a magnitude small compared to $\Delta_0$, and consequently that the susceptibility grows exponentially with the radius of the grain, according to Eqs. \ref{coefficients} and \ref{suscept}.  However, if $\sigma_R$ is sufficiently large, the optimal grains get to be large enough that  $R_{opt}-R_c \geq A \xi_0$ (where $A$ is of order 1), and consequently $|\Delta_i| \sim \Delta_0$.  In this limit,
the quantum dynamics of the order parameter is determined by
quantum fluctuations of the electromagnetic field,  and $\chi_{i}$
is determined by the normal state conductance of the metal, $G_{eff}$, as in Eqs. \ref{3D} and \ref{FeigLarkSusc}.
We can estimate the conditions for this by extrapolating Eq. \ref{optimalR} to the point
$R_{opt}-R_c = \xi_0$, from which we
deduce that the optimal puddles are ``large'' when $\sigma_R^2 > (\xi_0/\bar R)^2(1/Z''g)$.

In this limit, since $\chi$ grows with size of the grain only relatively weakly ($\log [\chi] \sim R^{D-2}$),
while the density of grains of large size falls exponentially with their volume, the density of optimal puddles is simply the density of grains with radius larger than  $R_0\equiv R_c + A\xi_0$,
\begin{equation}
N_{large} = \int_{R > R_0} dR P(R) \sim
N \exp\left[-\left(\bar{R}-R_{c})^{2}/\bar{R}^2\sigma^{2}_{R}\right)\right].
\end{equation}
With $N_{large}$ playing the role of $N_{opt}$, and with the expressions in Eq. \ref{3D} and \ref{FeigLarkSusc} for the susceptibility, the same analysis can be applied as in part {\bf B} to obtain $N_c$.  For instance, in 3D,
\begin{equation}
N_{c}\sim
R_c^{-3}\exp[-ZG^{eff}+(\bar{R}-R_{c})^{2}/2\bar R^2\sigma^{2}_{R}].
\label{N_disEM}
 \end{equation}
We would like to stress that Eq. \ref{N_disEM} holds only at exponentially small temperatures for which
$L_{T}\gg N_{c}^{-1/D}$. In the opposite limit, puddles with
$R_{i}>R_{o}$ are irrelevant, and the Cooper instability contribution represented by  Eq.\ref{suscept} dominates the physics of the phase transition, even for relatively large values of $\sigma_R$.

\begin{figure}
\begin{center}
\includegraphics[scale=0.5, bb=36 124 711 576]{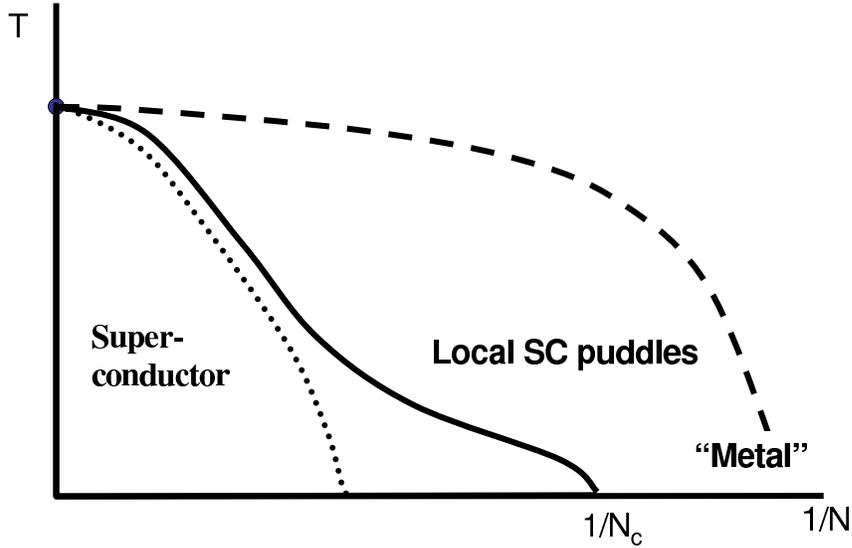}
\caption{ Schematic phase diagram for the case considered in Section IIB, in which a concentration $N$ of superconducting grains with a distribution of sizes is  embedded in a metallic host.
 The solid line represents a phase transition.
  The dashed line represents a crossover where a ``small fraction'' of the sample first supports a mean-field solution - although near this line, global phase coherence is destroyed by quantum and thermal phase fluctuations, measurable manifestations of local superconductivity onset below this line.  The dotted line represents the expected phase boundary in the Cooper limit, where mesoscopic fluctuations are ignored and a single uniform effective interaction between electrons is assumed as in Eq. \ref{barlambda}.  }
\label{fig1}
\end{center}
\end{figure}

\section{S-wave superconductor in a magnetic field}

\subsection{
A magnetic field perpendicular to a superconducting film}

A magnetic field, $H$, applied perpendicular to a metallic film ($D=2$) couples primarily to the electron's orbital motion.
In this case the Zeeman coupling of the magnetic field to the electron spin can be neglected.
To the extent that mesoscopic fluctuations of the order parameter
can be neglected, the problem of s-wave superconductivity in a
magnetic field was solved by Abrikosov and Gorkov. In
this
 approximation, at
$H^{(0)}_{c2} = \Phi_{o}/{\pi \xi^2_o}$, the  order parameter can be
represented  as a superposition of wave-functions in the first
Landau level,
\begin{equation}
\phi(r) = \frac{1}{\sqrt{2\pi L_Hd}}\exp(-r^2/L^{2}_{H}),
\end{equation}
where $\Phi_{o}=hc/2e$ is the flux quantum, $L_{H}=\sqrt{\Phi_0/2\pi H^{(0)}_{c2}}$
is the magnetic length.  (We consider the ``dirty limit'' $\xi_0 \gg l \gg k_F^{-1}$ where $\xi_{0}= \sqrt{D_{tr}/\Delta_{0}}$).
 Roughly speaking, the same form of the
wave-function applies even when mesoscopic fluctuations are taken
into account.  This simplifies the analysis in that it implies that,
 near the point of the transition, the puddles have a typical size,
$L_j\approx  L_{H}$.
However,  the critical magnetic field $H_{i}$ varies randomly as a function of position, so that,
in the notation of  Section I, we can identify
 \begin{equation}
 \gamma_i - \gamma \equiv \frac{(H_{i}-H)}{H^{(0)}_{c2}}.
\end{equation}

We will assume that  the distribution
of $H_{i}$ is approximately gaussian
\begin{equation}
P(H_{i})=\frac{1}{\sqrt{2\pi}\sigma_{H} \bar{H}_{c}}\exp[-\frac{(H_{i}-\bar H_c)^{2}}
{ 2\sigma_{H}^{2}\bar{H}_{c}^{2} }]
\label{PofH}
\end{equation}
and is characterizing  by the average $\bar H_{c}$ and a dimensionless
 variance $\sigma_{H}$. (This ignores  the existence of long, but for present purposes irrelevant tails of the distribution produced by
mesoscopic effects \cite{LernerKravtsov}.) We
assume that $\sigma_{H}\ll 1$, and thus that $H_{c2}^{(0)}\approx \bar H_c$. Generally, there are two
contributions to the variance $\sigma_{H}$: one contribution comes
from classical fluctuations in the strength of the local scattering
potential and one from non-local quantum ``interference'' effects,
\begin{equation}
\sigma_{H} = \sigma^{(int)} + \sigma^{(cl)}.
\end{equation}

The classical contribution is due to random fluctuations of the
concentration of impurities,
\begin{equation}
  \sigma^{(cl)}_{H}(L)\sim (\Lambda/L_{\bar H})
\label{sigma_cl}
\end{equation}
where $\Lambda\ll L_{\bar H}$ is the correlation length of the disorder potential.
The electron interference contribution is  \cite{spivakzhou}:
\begin{equation}
\sigma^{(int)}\sim
{1}/{G_{2D}}\ll 1.
 \label{sigmaHint}
\end{equation}
Note that, although the interference term is independent of puddle size, and hence is the larger term for big enough puddles, for large $G_{2D}$ there is a parametrically wide range of puddle sizes for which the simple statistical variations in impurity concentrations dominates the variance of local critical fields.

The configuration dependent (mesoscopic) variations in the Josephson coupling, $J_{ij}$  are more important here than in the previous example.
One can see this by noticing that
at large  $|{\bf r}_{i}-{\bf r}_{j}|$ (up to possible logarithmic corrections) the
 averages
\begin{eqnarray}
&& \overline {J_{ij}
}\sim \nu L_{H}^{2}d \frac{L^{2}_{H}}{|{\bf r}_{i}-{\bf
r}_{j}|^{2}}\exp\left[-\frac{|{\bf r}_{i}-{\bf
r}_{j}|}{L_{H}}-\frac{|{\bf r}_{i}-{\bf r}_{j}|}{L_{T}}\right] \nonumber \\
&&\overline{\tilde{J}_{ij}}\sim {G}_{eff}\frac{D_{tr}}{|{\bf r}_{i}-{\bf
r}_{j}|^{2}}\exp\left[-\frac{|{\bf r}_{i}-{\bf
r}_{j}|}{L_{H}}-\frac{|{\bf r}_{i}-{\bf r}_{j}|}{L_{T}}\right]
\label{JavH}
\end{eqnarray}
are much smaller than the variances
\begin{eqnarray}
&&\left[{\overline {|J_{ij}|^{2}} }\right]^{1/2} \sim  \left(\frac{L_{H}^{2}}{v_{F}l}
\right) \left(\frac{L_{H}}{|{\bf r}_{i}-{\bf
r}_{j}|}\right)^{2}
\exp
\left[-\frac{|{\bf r}_{i}-{\bf r}_{j}|}{L_{T}}\right]. \nonumber \\
&&\left[{\overline {|\tilde{J_{ij}}|^{2}} }\right]^{1/2} \sim   \frac{D_{tr}}{|{\bf r}_{i}-{\bf
r}_{j}|^{2}}
\exp
\left[-\frac{|{\bf r}_{i}-{\bf r}_{j}|}{L_{T}}\right].
\label{deltaJH}
\end{eqnarray}
where
$\overline {O}$ indicates the average of $O$
over realizations of the random scattering potential., and $v_{F}$ is the Fermi velocity.
As a consequence, at long distances the Josephson coupling in Eqs.~\ref{TDGL} and \ref{TDGL1}
is of the form \cite{spivakzhou}
\begin{eqnarray}
&&J_{ij}\sim F_{ij}\nu L_{H}^{2}d \frac{L_{H}^{2}}{|{\bf r}_i-{\bf r}_j|^{2}}\exp\left[-\frac{|{\bf r}_i-{\bf r}_j|}{L_{T}}\right]\nonumber \\
&&\tilde{J}_{ij}\sim \tilde{F}_{ij}\frac{D_{tr}}{|{\bf r}_i-{\bf r}_j|^{2}}\exp\left[-\frac{|{\bf r}_i-{\bf r}_j|}{L_{T}}\right]
\label{J_ijpm2}
\end{eqnarray}
where $F_{ij}=F_{ij}({\bf r}_{i},{\bf r}_{j})$, and $\tilde{F}_{ij}=\tilde{F}_{ij}({\bf r}_{i}, {\bf r}_{j})$ are  dimensionless functions which varies randomly in phase,  and $\overline{|\tilde{F}_{ij}|^2}\sim \overline{|F_{ij}|^2} \sim 1$.

To find the critical magnetic field $H_{c}$, we employ the same optimization procedure
we used in the Section 2. We introduce an interval in the space of $H_{i}$ which is centered at
$H_{opt}$ with width $\Delta H\sim \sigma_{H}\bar H_c$. We will see
that for sufficiently large values of $\sigma_{H}$, the distance between
the ``optimal puddles" is large enough that the Joshepson coupling between puddles is dominated by
the  mesoscopic contribution given by Eq.~\ref{J_ijpm2}.

Depending on the value of $G_{2D}=e^{2}\nu D_{tr}d$, the quantum dynamics of the order parameter
is determined by either the Cooper instability or the quantum fluctuations of the
electromagnetic field.

If $(H_{opt}-H)/H_{c2}^{(0)}\ll 1$, and hence $\Delta_{opt}\ll \Delta_{0}$, the quantum dynamics of the order parameter
is determined by the Cooper instability, so we can use Eqs. 2 and 5 to obtain

\begin{equation}
X_{opt}\sim a_{H}
\exp\left [Z'' g\left(\frac{H_{opt}-H}{\bar H_{c} }\right)-\frac{(H_{opt}-\bar{H}_{c})^{2}}
{\sigma^{2}_{H}\bar{H}_{c}^{2}}\right] \,\,\,\,\, {\rm at} \,\,\,\,\,\ T=0.
\label{XoptH}
\end{equation}
Here $a_{H}\approx \Delta_{0}/E_{F}$, and $g=\nu \Delta_{0}\xi_{0}^{2}d \approx G_{2D}$.  As before, $H_{opt}$ is the value which maximizes this expression, and the true critical field is then determined as the value of $H=H_c$ at which $X_{opt}=1$:
\begin{equation}
H_{opt} =\bar H_c\left[1 +(ZG_{2D}/2)
{\sigma_H^2}\right]
 ; \ \ H_c = \bar H_c\left[1 + (ZG_{2D}/4)
 \sigma_H^2\right].
\label{Hcweak}
\end{equation}
These expressions are self-consistent so long as
\begin{equation}
 \frac 2 {\sqrt{ZG_{2D}}} \gg \sigma_H \gg
 \frac 2 {ZG_{2D}}
\end{equation}
where the first inequality guarantees that  $(H_c-\bar H_{c})/\bar H_{c} \ll 1$ and the second that the optimal puddles are dilute.
The expression for $H_{c}$ in Eq. \ref{Hcweak}   has been obtained in Ref. \cite{LarkinGalitskii} by a different method.

In any puddle for which $(H_{i}-H)/H_{c2}^{(0)} \gtrsim 1$,
 the superconducting order is ``fully developed,''  $\Delta_{i}\approx \Delta_{0}$, so the dynamics of the order parameter
is determined by the quantum fluctuations of the electromagnetic field. In this case, the susceptibility of the puddle depends only on $G_{2D}$ as in Eq.  \ref{FeigLarkSusc}, and is independent of $H_i$.  Since the probability of finding such a puddle decreases with increasing $H_i$, the optimal puddles of this sort are those with $H_i \approx H_{opt}\approx H$.  The corresponding dimensionless coupling between these puddles is thus
\begin{equation}
X_{opt}\sim
\exp\left [2Z^\prime\sqrt{G_{2D}}-\frac{(H_{opt}-\bar{H}_{c})^{2}}
{\sigma^{2}_{H}\bar{H}_{c}^{2}}\right] .
\label{XoptH1}
\end{equation}
These puddles are always dilute compared to their size so long as $[H-\bar H_c] >\sigma_H\bar H_c$.
The critical value of $H$, determined by the condition $X_{opt}\sim 1$, is
\begin{equation}
H_c =\bar H_c (1+
\sqrt{2} \sigma_H  G_{2D}^{1/4})\,\,\,\,\, {\rm at} \ \ \ \ \ T=0.
\label{Hcfull}
\end{equation}

The issue of whether the global superconducting properties are dominated by weakly superconducting or ``fully developed'' puddles is settled by determining which of the expressions for $X_{opt}$ in Eqs. \ref{XoptH} and \ref{XoptH1} give the largest value.  In particular, the critical field is determined by the larger of the values given by Eqs. \ref{Hcweak} and \ref{Hcfull}.

Since in both these cases, the puddles are dilute, the phase of the Josephson couplings between optimal puddles is random.
Thus, the $T=0$ ordered state is glassy, and is an example of a ``gauge-glass.''  While this phase does not have long-range order, it supports a non-zero Edwards-Anderson order and is generally thought to have zero resistance.\cite{gaugeglass,ffh}

 In 2D, however, there is no ordered state at non-zero $T$, so there are only crossovers as a function of $H$ and $T$.  There is a characteristic field  $H^\star(T)$,  such that for $H < H^\star$, the coupling between optimal puddles is large (in magnitude) compared to the temperature;  here, the resistivity is, presumably, due to some form of variable-range-hopping of vortices, and so decreases exponentially with decreasing $T$.   Clearly, $H^\star(T) \to H_c$ as $T\to 0$.  It is only weakly $T$ dependent at low $T$, but in the temperature range such that $L_T$ is smaller than the typical spacing between optimal puddles, but large compared to the puddle diameter, $H^{\star}$ is determined by the condition that the concentration of superconducting puddles must exceed $L_T^{-2}$;  this leads to the unusual $T$ dependence:
\begin{equation}
H^\star(T)\sim
\bar H_{c}\left[1+2\sigma_{H}  \ln^{1/2}\left({L_{T}}/{L_{H}}\right) \right].
\label{HparT}
\end{equation}
The resulting schematic phase diagram is shown in Fig.2.  Note that
there are a series of additional crossovers that we have not addressed here, and which are not shown in the figure, which occur at fields greater than $H^\star$.    These crossovers characterize various energy scales in the anomalous metallic phase proximate to the superconducting glass.  More of the physics of the anomalous metal will be addressed in future studies.

\subsection{The case of a parallel magnetic field}

We now consider the opposite limit, in which the coupling of an applied magnetic field to the electron spin (Zeeman coupling) is significant, and the coupling to the orbital motion of the electrons can be neglected.  To a good approximation, this can be realized in a thin film of an
s-superconducting metal
in which the superconductivity is destroyed by an in-plane magnetic field $H_{\|}$.
The
situation in this case critically depends on the value of the
parameter $\Delta_{0}\tau_{so}$ (or, in other words, on the atomic weight of the metal) where
$1/\tau_{so}$ is the spin-orbit scattering rate.
In the case of relatively strong spin-orbit coupling, $\Delta_{0}\tau_{so}\ll 1$, on mean field level and
in the absence of mesocopic fluctuations, the transition is second
order. In this case, the effect of mesoscopic fluctuations on the character
of the transition is
qualitatively similar to the transition in the perpendicular
magnetic field considered in the previous  subsection.
In the case $\Delta_{0}\tau_{so}\ll 1$, however, the situation is very
different because on the mean field level the transition is
first order.  In this section, we will consider this case, and to simplify the discussion, we will consider it in the limit of zero spin-orbit coupling, $\Delta_{0}\tau_{so}\to \infty$.
In this limit,
the microscopic physics responsible for the
emergence of a puddle state is quite different,
and in particular the energy associated with the
formation of puddles is larger than in  the previous examples. Therefore the  various
manifestations of the physics, and especially the glassy character
of the phase at intermediate fields, is more robust.

If  mesoscopic fluctuations are ignored, the zero temperature transition is first order, with
 a discontinuous jump
in the spin-magnetization density from $m=0$ (in the superconducting state) to $m=\chi_{sp} H_{\|}$ in the metallic state, where $\chi_{sp}=\nu\hbar \mu_B^2$ is the normal-state spin (Pauli) susceptibility.  The critical field is given by the
well-known Chandrasekar-Clogston limit,
\begin{equation}
H^{(0)}_{c\|}=
{\Delta_{0}}/{\hbar \mu_{B}}.
\label{Ch}
\end{equation}
(In the disorder free case, there might appear a narrow regime of fields near $H^{(0)}_{c\|}$ in which a partially polarized
Fulde-Ferrel-Larkin-Ovchinikov  (FFLO) state occurs, but this is not relevant in the  case $l\ll \xi_{0}$, considered here.)

In disordered systems the critical magnetic field $H_{c\|}({\bf r})$
 exhibits spatial fluctuations. In the absence of spin-orbit scattering
 the value of $H^{(0)}_{c\|}$
given by Eq. \ref{Ch} is independent of $l$, which means that there is no
classical contribution to the dimensionless variance $\sigma_{\|}$, which is, instead,
determined entirely by mesoscopic interference
effects.  Thus, the  dimensionless variance is \cite{ZhouSpivalPar}
\begin{equation}
\sigma_{\|} \equiv  \frac{\left[
\overline{ \left( H_{c\|}-\bar H_{c\|}\right)^{2}}\right]^{1/2}
}
{\bar H_{c\|}}
\approx \frac{1}{G_{2D}}.
\label{varHpar}
\end{equation}
 and the correlation length of $H_{c\|}({\bf r})$ is of order $\xi_{0}$.

 According to general theorems \cite{Imry}, in 2D quenched disorder
 destroys first order transitions.  Rather, near a putative first order transition, a domain structure occurs with a characteristic domain size $L_{dom}$ which is exponentially large in the small disorder limit, $L_{dom} \sim \exp[ Z^{\prime\prime}(1/\sigma_{\|})^2]$.   In some cases, the putative first order transition is simply smeared and replaced by a crossover which becomes increasingly sharp as the disorder becomes weaker.  However, in the present case, since there is  clearly a superconducting phase for small enough $H_{\|}$ and (as we shall confirm) a non-superconducting phase for large enough $H_{\|}$, there must still be a sharp, continuous quantum phase transition at a shifted critical field, $H_{c\|}$.

 Specifically, in the present case, the superconducting domains are regions with magnetization  $m$ near 0, and with the local magnitude of the order-parameter, $|\Delta_i| \approx \Delta_0$, while the metallic regions have $m\approx \chi_{sp} H_{\|}$ and miniscule magnitude of the superconducting order.  The volume fraction of the two phases is a function of $H_{\|}$;  it is roughly a 50-50 mixture when $H_{\|}\approx H_{c\|}^{(0)}$, and the superconducting fraction decreases monotonically with increasing $H_{\|}$.   However, global phase coherence is not lost  at $H_{\|}\approx H_{c\|}^{(0)}$, where on the mean field level the superconducting fraction first fails to percolate.  Rather, as in the other problems we have examined,  it occurs when the Josephson coupling between superconducting regions becomes sufficiently weak, which in turn occurs when the superconducting fraction is small and the superconducting regions far separated.

Because the superconducting regions have a characteristic size large compared to $\xi_0$, and the magnitude of the order parameter is large, the dynamics of phase fluctuations is determined by electric field fluctuations, and consequently (according to Eq.~\ref{FeigLarkSusc})
$\chi_i \approx \Delta_{0}\exp\left[Z^\prime\sqrt{G_{2D}}\right]. $

To determine the distribution of Josephson couplings, we note that
in an SNS junction, when the normal part of the junction is
partially spin polarized, \cite{buzdin} the
Josephson coupling oscillates in sign as a function the coordinates.  Specifically, at $T=0$,
\begin{eqnarray}
&& \overline{\tilde{J}({\bf r,r'})}\sim \frac{G_{2D} D_{tr}}{|{\bf
r-r'}|^{2}}\exp(-\frac{|{\bf r-r'}|}{L_{H\|}})\cos(\frac{{\bf
|r-r'}|}{L_{H\|}}),
\nonumber \\
&&\left[\ \overline{\left | \tilde{J}({\bf r,r'}) \right |^2}\ \right]^{1/2}\sim \frac{D_{tr}}{e|{\bf
r-r'}|^{2}}
\label{J(r,r)}
\\
&&\tilde{J}({\bf r,r'})\sim F({\bf r,r'}) \frac{G_{2D} D_{tr}}{|{\bf
r-r'}|^{2}}\cos(\frac{{\bf
|r-r'}|}{L_{H\|}}) \label{J(r,r)1},
\nonumber
\end{eqnarray}
where $L_{H\|}=\sqrt{D_{tr}/\mu H_{\|}}$, and  $F(\bf r,r')$ is a  sample specific function $(|F|\sim 1$) which has random variations both in modulus, and in sign.

The mesoscopic fluctuations of $\tilde J$ again dominate the average at
 distances large compared to $L_{H\|}$. Thus we can estimate the critical magnetic field $H_{c\|}$ at which the zero temperature phase transition
to the metallic phase takes place. For $H_{\|} > \bar H_{c\|}$, the probability of finding a superconducting puddle is
$\sim \exp\left[-\left(H_{\|}-\bar H_{c\|}\right)^{2}/2\sigma^{2}_{\|}\bar H_{c\|}^{2}\right]$.
 As a result, following the same line of reasoning as in previous sections,
\begin{equation}
\frac{H_{c\|}-\bar H_{c\|}}{\bar H_{c\|}}\approx \sqrt{Z^\prime} \sigma_{\|} G^{1/4}_{2D}
\end{equation}.

\begin{figure}
\begin{center}
\includegraphics[scale=0.5, bb=36 124 710 576]{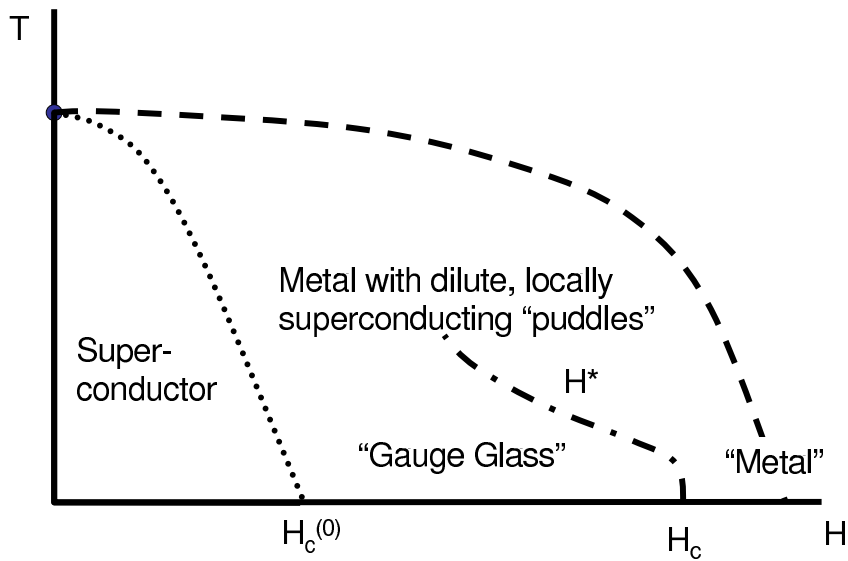}
\caption{Schematic phase diagram for
the cases considered in Sections IIIA and IIIB  in which superconductivity in a thin film of an s-wave superconductor  is destroyed, respectively, by application of a perpendicular or a parallel magnetic field.
The crossover scale, $H^\star$, (indicated by the dot-dashed line) is described in the text.  The ``gauge glass"
refers to a zero temperature phase in which the resistance vanishes as $T\to 0$, but for which there is no finite temperature phase transition.   The mean-field phase boundary  has a continuous portion (indicated by the narrower dotted line) and a first order portion (indicated by the heavier line) separated by a tricritical point (indicated by the solid circle).}
\label{fig1}
\end{center}
\end{figure}

Notice  that the average Josephson coupling in Eq.~\ref{J(r,r)}, itself, oscillates in sign, so even when the superconducting puddles are closely spaced, the Josephson couplings are random in sign.
The resulting glassiness of the superconducting state is
 more robust  than in the case of the
perpendicular magnetic field because of the large size of the superconducting domains, the fact that the magnitude of the  order parameter is fully developed within each domain, and due to the fact that the randomness in sign is not solely a long distance subtlety. In the absence of spin-orbit coupling, the magnetic field does not couple directly to any orbital degrees of freedom, and hence the glass phase can be precisely characterized
as in an XY spin-glass,
in which the ordered state supports spontaneously generated equilibrium orbital currents.

In the presence of spin-orbit coupling, a sharp definition of the glass phase based on the presence of orbital currents is not possible;  even a normal disordered metal phase will support small orbital currents under these circumstances.  As in the previous section, the superconducting coherent state near the critical field is some form of a gauge glass.

\section{Destruction of d-wave superconductivity by disorder}

\par We shall now consider the case in which, in the zero disorder limit, the superconducting state is a BCS state with d-wave symmetry
due to a weak attractive interaction in the d-wave particle-particle channel.

In the d-wave case, it is necessary to explicitly treat the dependence of the superconducting order parameter on the relative coordinate.
Specifically, in the absence of disorder and in a bulk sample,
 $\Delta({\bf r,r'})=\Delta^{(d)}\kappa_{d}^{(0)}({\bf r}-{\bf r}')$, where $\kappa_{d}^{(0)}({\bf r})$ changes sign under rotation by $\pi/2$, and is a short-ranged function, with  range, $b \ll \xi_0$, determined by the range of the effective attractive interaction.

 Although the ``d-wave'' notation is inherited from spectroscopic notation for an ``$l=2$" irreducible representation of the rotation group, in a crystal, it refers to an appropriate irreducible representation of the point group.  We will treat the case in which the point group has at least two distinct even parity one dimensional representations  - a trivial one and a non-trivial one.  For instance, in a tetragonal crystal, in addition to the trivial (s-wave) representation, there are three other even parity irreducible representations: a d$_{x^2-y^2}$-wave, a d$_{xy}$-wave, and a g-wave (which transforms like $(x^2-y^2)xy$).  We consider the case in which in the zero disorder limit, there is an effective attraction only for one of these representations, which we will call simply the ``d-wave.''

The most clear-cut manifestation of the d-wave nature of the ground state order parameter comes from
 ``phase sensitive'' measurements \cite{harlingen,Kirtley} of the symmetry of the order parameter.  Specifically, in a corner SQUID\cite{harlingen} of the sort described in Fig. 4, in which the external circle is a conventional s-wave superconducting wire, the ground state of the system will contain a half-flux quantum trapped  in the SQUID for the case in which the sample is a d-wave superconductor, and no flux if it is an s-wave superconductor.

 The fact that, in the absence of disorder,  $\Delta({\bf r},{\bf r}')$ changes sign
under rotation makes the system  much more sensitive to  the strength of the
 disorder than a conventional s-wave superconductor.
 We will characterize the disorder strength by the electron mean free path $l$  in the normal metal.
 What happens to the system in the presence of relatively strong disorder depends on the sign
 of the interaction in s-channel. If  the interaction in the s-wave channel is attractive but much weaker than the attraction in the d-wave channel,  then when the disorder is weak enough ($l > \xi_{0}$), the d-wave state dominates, but when the disorder strength increases enough to destroy the d-wave superconductivity ($l<\xi_{0}$), the system undergoes
 a phase transition to an s-wave state (See for example \cite{Vojta}). The s-wave state is
 destroyed
only when $k_{F}l\approx 1$. However,
in this article we consider the more interesting case in which the interaction in the s-channel is repulsive,  so when disorder suppresses d-wave superconductivity, it
  drives the system to a normal state when the mean free path is still relatively large, $k_{F}l \gg 1$.  This case may be relevant, for instance, to the destruction of superconductivity in the ``overdoped'' high temperature superconductors.

 In the (conventional)  approximation in which  spacial fluctuations
of the electron mean free path
are neglected, d-wave superconductivity is destroyed when  $l\sim l_0=1.78 \xi_o$.
Thus disordered d-wave superconductors are another example of a system which may have a quantum superconductor-metal
transition in a situation in which the conductance is large.
This case exhibits both similarities and differences with the  cases we have already considered in Sections 2 and 3.

In the presence of disorder, a material has no particular spatial symmetry, and so the order parameter cannot be said precisely to have any particular symmetry at all. Nevertheless, in bulk samples, symmetry is restored upon configuration averaging. It is therefore legitimate to ask questions concerning the global symmetry of the order parameter.
 Hence, we can ask whether
  $\overline{\Delta({\bf r},{\bf r}')}$, or
  $\overline{{\cal F}({\bf r},{\bf r}'})$  have d-wave or s-wave symmetry. Here   the overline stands for a configurational averaging, and
  ${\cal F}({\bf r,r'})\equiv {\cal F}({\bf r,r'}, t=t')$ is the anomalous Green function which is connected to $\Delta({\bf r},{\bf r}')$ by the interaction constant.

It is important to realize that it is possible (indeed, as we shall see, inevitable near criticality) to have a situation in which the local pairing is ``d-wave like'' and yet the global superconductivity has  s-wave symmetry.
 In fact we will show that there are at least two quantum phase transition as disorder increases: the transition from d-wave to s-wave global symmetry, and subsequent transition from  s-wave superconductor to the normal metal.  The  existence of the second (d-s) transition is the main difference with the
cases considered in previous sections.  (In fact, we consider it  likely that rather than a sharp d to s transition, there is an intermediate glass phase in which time reversal symmetry is broken and s and d-wave ordering coexist.  However, we have not fully explored this scenario.)

We
propose several different definitions of the global
symmetry of the order parameter:
 a) The best operational definition
 is provided by the result of a phase sensitive experiment, such as the corner SQUID experiment
 shown in Fig. 4.
 b) The quantity
 $\overline{\Delta({\bf r},{\bf r}'})$ can be characterized as having d-wave or s-wave symmetry.   It can also provide a definition of a state with coexisting order if it has mixed symmetry.
  c) A specific diagnostic for a globally s-wave component of the order parameter can be defined in terms of
  the
  local component of the anomalous Green function
  ${\cal F}({\bf r}={\bf r}')\equiv {\cal F}^{(s)}({\bf r})$.  If we define $P_{\pm}$ to be  the volume fraction
  of a sample
where $F^{(s)}({\bf r})$ has a positive or negative sign, respectively, then
the system has an s-wave component if
$(P^{+}-P^{-})\neq 0$.
These definitions are not equivalent under all circumstances.  However, for the purposes of  this article, we are not primarily interested in the most
general definition of the
global
symmetry of the superconducting state.
For the most part, we will deal with
  the interval of  parameters
  in which all these definitions are approximately interchangeable.

  \begin{figure}
\begin{center}
\includegraphics[scale=0.5, bb=36 101 733 576]{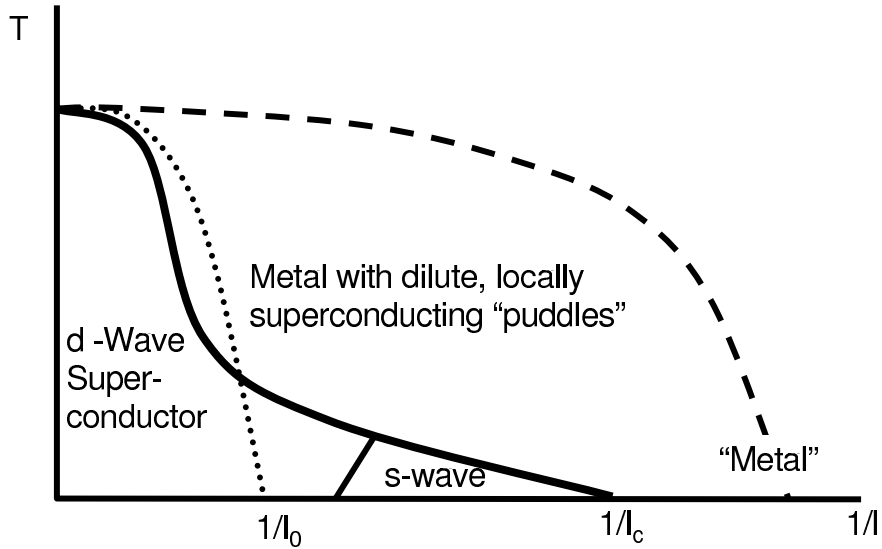}
\caption{Schematic phase diagram for the case considered in Section IV in which a BCS (weak-coupling) d-wave superconducting state (in 3D) is destroyed as a function of increasing disorder strength.  The dotted line represents a transition between d-wave superconuctor and normal metal calculated by the conventional theory.   In the presence of disorder, the labels ``s-wave'' and ``d-wave'' refer to the behavior of the system in a macroscopic phase sensitive measurement, as described in the text.  (The negative slope of the boundary which divides globally d- and s-wave global superconductors shown in the figure can be justified only in the case when the electron interaction in the s-channel is attractive, so that the entropy of the s-wave superconductor is smaller than the entropy of d-wave.  More generally, the slope of this boundary is to be determined.)}
\label{fig1}
\end{center}
\end{figure}

\subsubsection{The
d-wave to s-wave transition as a function of disorder}

This transition takes place in the region of concentrations where quantum fluctuations of the
order parameter can be neglected, and therefore it can be understood on the mean filed level.
As a warmup exercise, consider a cartoon picture of
  a system of superconducting puddles of a size large compared to $\xi_0$ and of a rectangular shape which are embedded in a noninteracting diffusive normal metal (See Fig. 5).
 The rectangles are identical, and they are oriented either vertically, or horizontally in  a random fashion. The order parameter inside the rectangles has d-wave symmetry, and the orientation of the gap nodes is assumed to be pinned by the crystalline anisotropy.

In a d-wave superconductor, in addition to an overall phase of the order parameter, there is an arbitrary sign associated with the internal structure of the pair wavefunction.  Specifically, we adopt a uniform phase convention such that when the phase of the order parameter $\phi_i= 0$, this implies $\Delta({\bf r}, {\bf r}')$ in puddle $i$ is real and has its positive lobes along the (appropriately defined) $y$ axis and its negative lobes along the $x$ axis.

 It is obvious that at a high concentration of puddles, the order parameter  in the ground state has global  d-wave symmetry (See Fig. 5a.). However at small  puddle concentrations, the situation is different.
  If the distances  between puddles $|{\bf r}_{i} - {\bf r}_{j}|\gg R$ are much larger than the characteristic size of the puddles, $R$, the  Josephson coupling between  puddles inevitably favors globally s-wave superconductivity, even though the order parameter on each puddle looks locally d-wave like. In this case the mean field exchange energy of the system has a form
 \begin{equation}
 E_{Jos}=\sum_{i\neq j} \eta_{i}\eta_{j}\tilde{J}^{(S)}_{ij}\cos(\phi_{i}-\phi_{j})
 \label{MFdsBigR}
 \end{equation}
 where $\eta_{i}=\pm 1$ are random numbers such that $\eta_i=1$ for a rectangle oriented in the x-direction and $\eta_{i}=-1$  for a y-directed rectangle.

Eq. ~\ref{MFdsBigR} represents the Mattis model which is well known  in the theory  of spin glasses \cite{mattis}.
   The ground state of this model corresponds to
\begin{equation}
\cos (\phi_{i})= - \eta_{i}.
\end{equation}
 Thus the distribution of $\exp(\phi_{i})$ between puddles looks completely random, as shown in Fig. 5b.
 However the system is not a glass because it's ground state has a hidden symmetry, which
 in the present problem corresponds to
 to a global s-symmetry of the order parameter according to any of our proposed definitions!

A qualitative explanation of this fact is as follows:
 The inter-puddle Josephson coupling originates from the proximity effect in the normal metal, which is characterized by the anomalous Green function ${\cal F}({\bf r, r'})$.
  Due to lack of   symmetry at the boundary of a puddle, an
 s-wave component ${\cal F}({\bf r}={\bf r}')={\cal F}^{(s)}({\bf r})\neq 0$ of the anomalous Green function is generated in the neighboring metal.  Specifically,
 at a normal metal-superconductor boundary, the sign of ${\cal F}^{(s)}({\bf r})$ is determined
by the sign of the d-wave order parameter in the {\bf k}-direction perpendicular to the boundary (See Fig. 5). (Thus
the sign of ${\cal F}^{(s)}({\bf r})$ changes along the boundary of a puddle.)
 On distances from the boundary larger than $l$, the anomalous Green function
  becomes isotropic. In other words, only the s-component $F^{(s)}({\bf r})$ survives elastic scattering.
   It is this component which penetrates through the metal
   and carries the Josephson current between puddles.
   At distances larger than the size of the puddle (but smaller than $|{\bf r}_{i}-{\bf r}_{j}|$) the quantity  ${\cal F}^{s}({\bf r})$ has a definite sign which is determined by an integral around the surface, which sign gives us the value of $\eta_{i}$.

\begin{figure}
\begin{center}
\includegraphics[scale=1, bb=36 358 445 576]{OSMFig4.eps}
\caption{A schematic picture of a phase sensitive ``corner SQUID'' experiment, introduced in Ref. \cite{harlingen}.  If the square piece of superconductor has  global d-wave superconductor symmetry, then there is a magnetic flux trapped in the ground state of the system.
Pluses and minuses inside rosettes indicate the sign of $\Delta({\bf k})$ as a function of the direction of ${\bf k}$ }
\label{fig2}
\end{center}
\end{figure}

\begin{figure}
\begin{center}
\includegraphics[scale=0.6, bb=36 78 740 576]{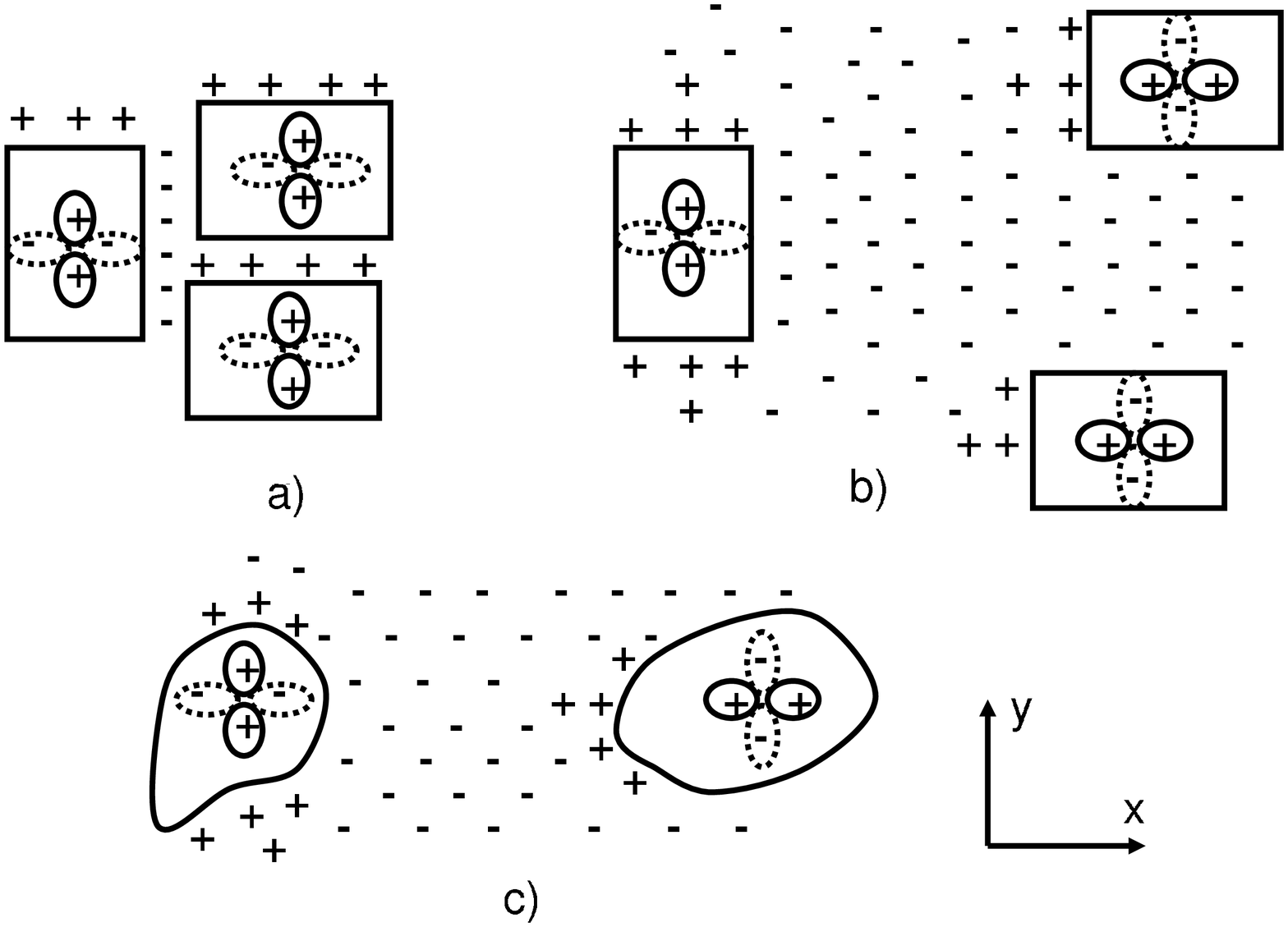}
\caption{A qualitative illustration of the global d-wave to s-wave transition. Solid lines represent boundaries of d-wave superconducting puddles embedded into a normal metal.
Pluses and minuses indicate the areas were the s-wave component of the anomalous Green function $F^{s}({\bf r}, {\bf r})$ is positive and negative respectively.  a) The ``cartoon'' case in which the concentration of  regular rectangular d-wave puddles
is large, so the system has d-wave global symmetry. b) The case when the concentration of d-wave puddles is small so the system has s-wave global symmetry. c) The case in which the concentration of  d-wave puddles is small and they have arbitrary shapes.  Here, they are shown embedded in a normal metal, and the system has global s-wave symmetry.}
\label{fig3}
\end{center}
\end{figure}

On intermediate distances, the situation is more complicated.
Areas with different signs of ${\cal F}^{(s)}({\bf r})$  mix in a random fashion.  We
argue that the most important
aspects of this complex situation can be  modelled by the following  effective Josephson energy:
\begin{equation}
E_{Jos}=\sum_{i\neq j} [\eta_{i}\eta_{j}\tilde{J}^{(s)}_{ij}+\tilde{J}^{(d)}_{ij}]\cos(\phi_{i}-\phi_{j})
\label{JoshepsonDS}
\end{equation}
where $\tilde{J}_{ij}^{(d)}$ characterizes the strength of the exchange interaction between the d-wave components
of the order parameter. Typically at small $|{\bf r}_{i}-{\bf r}_{j}|$,
$\tilde{J}_{ij}^{(d)}>\tilde{J}_{ij}^{(s)}$, but
at large $|{\bf r}_{i}-{\bf r}_{j}|$ the coupling strength $\tilde{J}_{ij}^{(s)}$ decays more slowly than $\tilde{J}_{ij}^{(d)}$.
Thus between the asymptotic d-wave dominated regime where the puddles are dense, and the s-wave dominated regime where they are dilute, it is likely that
there is an intermediate region in which the system will exhibit spin glass features and/or coexistence of d-wave and s-wave ordering. In this article, however, we will not
further explore this fascinating but complex aspect of this problem.

While the above discussion was based on a cartoon model with regularly shaped puddles, we would like to stress that  our conclusions do not rely on this.
In particular, as it is qualitatively illustrated in Fig. 5c,  Eq. \ref{MFdsBigR} holds at arbitrary shape of the puddles provided that the typical distance between them is larger than their characteristic size.

To quantify
our conclusions, we compute the Josephson coupling between a pair of far separated puddles in two extreme limits,  large puddles and small puddles:

If the size of the puddles is large enough, the Josephson coupling has to be obtained from the solution of
  the Usadel equations (See for example \cite{likharev})
for the configuration averaged anomalous Green function  $\overline{{\cal F}_{\epsilon}^{(s)}({\bf r})}\equiv -i\sin \theta(\epsilon, {\bf r})$
 in the metal,
\begin{equation}
\frac{D_{tr}}{2}\partial^{2}_{{\bf r}}\theta(\epsilon, {\bf r})+i\epsilon \sin \theta(\epsilon, {\bf r})=0.
\label{UnadelEq}
\end{equation}
Here ${\cal F}_{\epsilon}^{(s)}({\bf r})$ is the Fourier  transform of ${\cal F}^{(s)}({\bf r},t-t')$, and
the symbol $\overline{0}$ indicates averaging over the random scattering potential between the puddles at given shape of the puddles.
The boundary conditions for Eq.~\ref{UnadelEq} at the normal-superconductor surface
have been derived in Ref. \cite{Nazarov}. They are valid as long as the size of the puddles is large and the Andreev reflection on the puddles is effective.

Since the relevant energy for computing the Josephson coupling, $\epsilon\approx D_{tr}/|{\bf r}- {\bf r}_j|^{2}$, is much smaller than the value of the order parameter in the puddles,
  the boundary condition for $\theta({\bf r},\epsilon)$ is independent of $\epsilon$, and  depends only on the angle
between the unit vector parallel to the direction of a gap node, ${\bf \hat n}_{\Delta}$, and a unit vector, ${\bf \hat n}({\bf r})$, normal to the boundary at  point ${\bf r}$ at the surface, :
\begin{equation}
\theta_{s}(\epsilon, {\bf r})=f[\alpha({\bf r})], \ \ \  \sin[\alpha({\bf r})]\equiv{\bf \hat n}({\bf r})\cdot{\bf \hat n}_{\Delta}.
\label{boundCond}
\end{equation}
Here $f(\alpha)$ is a smooth, approximately odd and periodic function, $f(\alpha)\approx -f(-\alpha) $, $f(\alpha) \approx f(\alpha+\pi)$,  which grows from $f(\alpha)\approx 0$ at $\alpha=0$, to $f(\alpha)\approx \pm \zeta$ for  $\alpha=\pi/4$. Here $\zeta \sim 1$.

Solving Eq. \ref{UnadelEq} with  the boundary conditions Eq. \ref{boundCond}  and using the standard procedure of calculation of the Josephson energy we get
\begin{equation}
\eta_{i} = {\rm sign}\left\{ \int_{i} d s f(\alpha) \right\}
\end{equation}
where the integral is taken over the surface of the $i_{th}$ puddle.
Moreover, the value of $\tilde{J}_{ij}^{(s)}$
in Eq.~\ref{MFdsBigR} turns out to be of the same order as  in Eq.~\ref{TDGL1}.
It is this long range nature of the decay which ensures the existence of the phase in which
the puddles separates by a large distance
and the system has global s-wave phase coherence.

In the second case, the Andreev reflection is ineffective and
the interactions between puddles can be computed in perturbation theory.
 Thus we can write an analog of Eq.~\ref{MFdsBigR}.
\begin{equation}
E_{Jos} = -\sum_{i\neq j}  \eta'_{i}\eta'_{j} [J^{(s)}_{ij} \Delta_i^\star\Delta_j +
{\rm c.c.}]
\label{S_J}
\end{equation}
where,
$\Delta_i$ is (up to a sign, $\eta'_j=\pm 1$) the average of the order parameter over puddle $i$,
\begin{equation}
\Delta_{i}\equiv \eta_i\int_{{\rm puddle} \ i}\Delta({\bf r}, {\bf r}') \frac {d {\bf r} d {\bf r}'} { V_{i}^{2}}  ,
\label{Delta_s}
\end{equation}
and $\eta'_i$ is a random variable that we have introduced in Eq.~\ref{S_J},   (and then cancelled in the definition of $\Delta_i$).
The strength of the Josephson interaction between s-component of the order parameter in puddles characterized
by $J^{(s)}$ is (up to a numerical factor smaller than one) of order of Eq.~\ref{J_ijpm}.
Again, we have neglected in Eq.~\ref{S_J} the interactions between the d-wave components of the order parameter, since they fall faster with separation between puddles.
Notice that in the fine tuned case of a fully symmetric puddle, $\Delta_i=0$ due to the d-wave symmetry.

The important point is that, in both the small and large puddle limits,
  Eqs.~\ref{S_J}, and \ref{MFdsBigR}  yield the same qualitative picture:
at large inter-puddle distances the
Josephson coupling favors s-wave symmetry.
It now remains to show that, near the point of quantum SMT,
the distance between optimal puddles is indeed much larger than their size.

\subsubsection{ Globally s-wave
superconductor to metal transition}

The quantum transition between a globally s-wave superconducting  state and the metal
does not differ in a crucial way from the transition which has been considered in Section 2.
For reasons that should by now be familiar, near the critical value of the disorder
the spacial dependence of the order parameter can be visualized as defining a system of far separated superconducting
puddles with anomalously large values of the order parameter separated by large areas of the normal metal.
In particular, this results in a smaller value of $l_c$ (larger critical magnitude of the disorder strength) for the destruction of superconductivity than $l_{co}$ which is given by the conventional theory.
The difference between the present problem and that treated in Section 2 is that,  to identify a set of optimal puddles,   one has  to characterize them by two generally independent parameters: the size of the puddles and the value of the s-component of the order parameter associated with an optimal  puddle.

As we have seen in previous sections, depending on whether the Andreev reflection from the superconductor-metal boundary is effective or not, two scenarios are possible. In the first case the amplitude of the Josephson energy is independent of the value of the order parameter in the puddles, and is given by Eq.~\ref{J_ijpm1}. In the second case the amplitude of Josephson energy is
given by Eq. \ref{J_ijpm}.
The susceptibility of an isolated puddle does not depend in any essential way on the symmetry of the superconducting order parameter, and as such, the analysis follows along identical lines as for the case of s-wave puddles, discussed in Sec. II.
Thus, for small puddles, the susceptibility is determined by the Cooper instability Eq.~\ref{suscept}, while  puddles of radius $R$ which is large compared to the coherence length, the susceptibility  is determined, as in Eq.~\ref{3D}, by the effective conductance, $G^{eff} \sim  R^{D-2}$, where $R$ is the radius of the puddle.

Let us consider a situation where the mean free path $l({\bf r})$ is a random quantity which  exhibits classical spatial fluctuations with a Gaussian distribution
\begin{equation}
P(l)=\frac{1}{\sqrt{2\pi}\sigma_{l}\bar{l}}\exp[-\frac{(l-\bar{l})^{2}}{2\sigma^{2}_{l}
\bar l^{2}}]
\end{equation}
characterized by the average ${\bar l}$, a dimensionless
variance $\sigma_{l}$, and a correlation length, $\Lambda$.

To be concrete we consider the 3D case, and neglect the mesoscopic fluctuations of the mean free path of an interference nature.
We also assume that the conductance of the metal is isotropic.

The size of the optimal puddles is readily seen to be of order the coherence length
$R_{opt}\sim \xi_{opt} \sim \xi \sim \xi_{0}\sqrt{l_{0}/(l_{opt}-l_{0})}$ , and therefore
the susceptibility of the puddles is given
by the large puddle result in Eq.  \ref{3D}.
(The self-consistency of this assumption can be checked  starting with the assumption that the small-puddle expression in
Eq. \ref{suscept} can be used, and then determining the optimal puddle size -- this procedure
  leads to the inconsistent conclusion that the optimal puddles are arbitrarily large.)

The effective conductance that determines the susceptibility is the
conductance of a region of characteristic linear dimension $\xi_{opt}$;  this, in turn, depends on the local  value of the mean-free path as
$G_{opt} = G_{\xi_{0}}\sqrt{l_0/(l_{opt}-l_{0})}$, where
$G_{\xi_{o}}$ is a conductance of
a region of size $\xi_{0}$.
The probability to find a puddle of this size is determined by
the variance of the mean free path averaged over it's volume, which is of order  $\sigma_{l}(\Lambda/\xi)^{3/2}$.
Thus we have
\begin{equation}
X_{opt}\sim \exp\left [2ZG_{\xi_0}\frac {l_0^{1/2}} { (l_{opt}-l_0)^{1/2}}-\frac{(l_{opt}-\bar{l})^{2}}{\sigma_{l}^{2}(l_{opt}-
l_0)^{3/2}l_0^{1/2}}\left(\frac{\xi_{0}}{\Lambda}\right)^{3}\right].
\label{X_0^D}
\end{equation}
As usual, $l_{opt}$ is the value which maximizes Eq.~\ref{X_0^D}, and the critical disorder, $\bar l=l_c$, is obtained by equating the result to unity.  Although $ l_c > l_0$, as long as $\sigma_l$ is sufficiently small, $(l_{0}-\bar{l}_{c})/\bar{l} \ll 1$.  In this limit, the result of this procedure can be expressed
as
\begin{equation}
(l_{0}-l_{c})/l_0 \sim
 2Z G_{\xi_{0}}\sigma_{l}^{2}\left(\frac
 {\Lambda}{\xi_{0}}\right)^{3}
\end{equation}

 Near the critical point $\bar{l}=l_{c}$, the distance between the optimal puddles $R_{opt}\exp[G^{1/2}_{\xi_o}(\xi_{0}/\Lambda)^{3/2}/\sigma_{l} ]\gg R_{opt}$  is exponentially large.
 The assumption that the distance between optimal puddles is large which we made in the previous subsection is, thus, justified
for $\bar{l}$ near $l_{c}$.

The  temperature dependence of the critical value of the mean-free path, $l_{c}(T)$, can be obtained from similar considerations to those used to determine  $N_{c}(T)$ in Section 2.

The resulting phase diagram for a d-wave superconductor in the presence of quenched disorder is shown schematically in Fig. 3.
At large disorder ($l<l_{c}$) the system is in the normal metal phase.
At $l_{0}>l_{c}$ the system in a state with a
dominant s-wave component of the order parameter, and a d-wave component, whose sign is locally slaved to the s-wave component, in a way  which varies randomly in space.
At still larger values of $l>l_{0}$ ,  there is a
dominantly d-wave state, in which the s-component has a sign is locally slaved to the d-component and varies randomly in space.

We would like to note that both d-wave and s-wave superconducting phase are expected to exhibit glassy behavior associated with rare regions where the uniform phased order is strongly frustrated.  We consider the existence of a glass phase which spontaneously breaks time reversal symmetry likely but not proven, so we have not shown it in the phase diagram in Fig. 3.

\section{Discussion}

In this article we have focussed attention on several systems in which,
at the point of the quantum superconductor-metal transition,
the conductivity is still large compared to the quantum of
conductance, and hence localization effects are unimportant.
The key general aspects of this transition are:   1)  The $T=0$ transition occurs at a point at which the superconducting order parameter is small in most of the sample, other than in a dilute set of locally superconducting puddles.  2)  The transition is triggered by the quantum fluctuations of the phase of the order parameter on these puddles, whose quantum dynamics can either be governed by the dynamics of the Cooper instability (when the optimal puddles are small) or the electric field fluctuations (when the optimal puddles are large).  3)  While there is probably a small quantum critical regime at exceedingly low temperatures and very close to the quantum critical point where the physics is universal and can be described by a suitable quantum critical scaling theory, there is a much larger regime where quantum phase fluctuations dominate the physics, but the long-distance properties are more properly described by the thermally truncated percolation of phase coherence between puddles.

We have  ignored the fundamental, but for our purposes purely academic
  question of  electron localization in highly conducting samples.
However, there remains the issue of the ultimate fate of the ``metallic phase'' in the true $T\to 0$ limit.  In 3D this is, presumably,
 not an issue, but in 2D, where all single particle states are localized, this is an important point of principle.  For $G_{2D}\gg 1$ the temperature below which interference effects are relevant is exponentially small, so the issue is not of practical relevance. Still even the point of principle is interesting, and, in our opinion, unresolved partially  because  calculations
of weak localization corrections close to the point of the metal-superconductor transition, where the conductivity is much larger
than the Drude value, remains a challenge.

\subsubsection{Previous studies of the problem }

There have, of course, been many
theoretical studies of the quantum phase transition from the superconducting to the non-superconducting state in the presence of quenched disorder.  Many of these studies concerned a scenario
in which there is a direct superconductor to insulator transition.
Under circumstances in which, near criticality, $k_F l \approx 1$, it is conceivable that there is a direct superconductor to insulator transition.  Moreover, in common with the superconductor to metal transition considered by us, near such a superconductor to insulator transition
the electron wave functions become strongly non-uniform, and exhibit fractal features \cite{KravtsovAronov}. Therefore it is not surprising that the order parameter at the point of the transition also non-uniform \cite{FeigelmanKr,trivedi}.
Despite the similarity of this aspect of the two transitions, the superconductor to insulator  transition differs from the superconductor to metal transition in significant ways, and is outside the scope of the present paper.

There have also been many previous studies of the superconductor to metal transition, starting with the mean-field studies of Abrikosov and Gorkov of the transition in a magnetic field.
In this context,
 we would like to mention the paper Ref.
\cite{Finkelstein} where a renormalization group approach was proposed for disordered 2D s-wave superconductors in the absence of magnetic field.   The conclusion reached in this article bears  some similarity
to ours: The superconductivity is quenched under conditions such that $G_{2D} > 1$.
In the framework of the analysis in Ref. \onlinecite{Finkelstein}, the reason is that the
diagrams responsible for suppression of $T_{c}$ by
fluctuations of the phase of the order parameter are proportional to
$\ln^{3}(L_{T}/l)$, while the weak localization corrections to
the conductivity are only proportional to $\ln (k_{F}L_T) $.  The theory presented in this work
was also supported by a comparison between theory and the experiments of Ref.  \onlinecite{MBisley}.
The corrections of order $\ln^{3}(L_{T}/l)$ in Ref.\cite{Finkelstein}  are of the
similar physical origin as the power law in Eq. \ref{decay}.  However, there are significance differences in our analysis, the most important of which is that
in the
situations we have considered, the existence of a superconductor-metal transition is generically
unconnected with interference effects governed by the parameter $k_{F}l$.
Therefore
the quantum superconductor to insulator transition and
2D localization
can be treated separately from each other. As a result, the effects we have considered are much larger than those considered in Ref. \onlinecite{Finkelstein}.

Pioneering studies of the quantum d-wave superconductor (or more exotic superconductor) to metal transition
were carried out in Refs. \cite{Herbut,Galitski2}, which employed standard diagramatic techniques and  the replica trick to implement the disorder averaging.  Our results
differ significantly from those in Refs. \cite{Herbut,Galitski2},
principally due to the fact that these earlier studies did not
 account
 for existence and crucial role of rare suprconducting puddles near the transition.

\subsection{Experiments on quantum superconductor-metal transitions}
\label{experiments}

There have been a vast number of experiments on the destruction of superconductivity in thin-film systems
 - too many for us to comment on here. For example there are many experiments (for a review see, for example, Ref. \cite{goldman}) in which
 s-superconductivity in films is destroyed by disorder in the absence of a magnetic field, and the transition takes place at $G_{eff}\sim 1$. Since our theory is developed in the case $G_{eff}\gg 1$, the present theory has no direct relevance to these experiments.
 We will only discuss  experiments in which $G_{eff}\gg 1$, and even here, only a very small subset of them.

Before discussing the relation between our results and experiments we must address
a question of terminology. An unambiguous distinction between
metallic and insulating states can be made only
in the limit $T\to 0$.
The metallic state has finite
resistance in this limit, while the insulating state has infinite resistance.  (The superconducting state, of course, has  zero resistance.)
The complication concerns the way in which finite temperature data is extrapolated to the $T\to 0$ limit.

One relatively widely used criterion is to study the sign of the dimensionless quantity
\begin{equation}
R_T\equiv d\log[\rho]/d\log[T]
\end{equation}
 at the lowest accessible temperatures, and to identify the insulating state with $R_T < 0$, a superconducting state with $R_T >0$, and a metallic state with $R_T\approx 0$.  Clearly, in the extremes, this is a sensible criterion, since in order for the resistivity to either vanish or diverge in the $T\to 0$ limit, $R_T$ must  have the stated sign.  The problem comes when $R_T$ in the accessible range of temperatures is relatively small in magnitude.  There are certainly well documented ways for a metal to exhibit $R_T < 0$ (for instance, in the Kondo effect), so an observation of $R_T <0$ cannot be safely taken as evidence that $\rho$ will diverge as $T\to 0$.
Of course, it is also common for metals to have $R_T >0$, so  this observation, by itself, cannot be taken as a sure indication of a zero temperature superconducting state.
It is
always
also important to pay attention to the absolute magnitude of the resistance in such discussions.
Metals at low temperatures typically
have resistances smaller than
the quantum of resistance, while insulators, at low enough temperatures, always have resistances large compared to this value.

\subsubsection{Transition in a perpendicular magnetic field in films with large $G_{2D}$}

When highly conducting films of a superconducting metal, such as MoGe, are subjected to a perpendicular field, two asymptotic behaviors are expected, and observed as a function of $H$:  For small enough $H$, the resistance tends \cite{MasonKapitulnik} to arbitrarily small values at low $T$,  which can be interpreted as a superconducting state.  At large enough $H$, the resistance tends to a roughly temperature and $H$ independent value, $\rho=\rho_D$, which can be interpreted as the ``normal'' state (Drude) value.  ($\rho_D$ is expected to be almost field independent so long as $r_{c}l\ll 1$  where $r_{c}(H)$ is the cyclotron radius.)

It has been observed in \cite{MasonKapitulnik}, and latter in \cite{Yoon}, that at small $T$
 there is a large interval of magnetic fields where the resistance is independent of $T$, and can be up to four
orders of magnitude smaller than $\rho_{D}$.

Although we have not calculated the conductance, and so cannot propose direct comparisons between theory and experiment,  we believe that the significant enhancement of the conductance
takes place in the interval of magnetic fields in which somewhat isolated puddles of the sample have a local value of $H_{c2} > H$.  (It is not completely clear, to us, whether the observed behaviors should be interpreted as the finite $T$ behavior of a system in the range of fields, $H_{c2}^{(0)}<H<H_{c}$, where the system forms a gauge glass phase in the $T\to 0$ limit, or whether it should be interpreted in terms of the anomalous metallic phase for $H \gtrsim H_c$, where even at $T=0$ there is no global phase coherence.)
In any case, to  explain the broad range of $H$ over which significant superconducting fluctuations occur,
we must assume that $\sigma_{H}>\sqrt{G_{2D}}$ in Eq. \ref{PofH}.

In this context we  would like to mention an interesting  phenomenological observation made recently by Steiner {\it et al} \cite{SteinerBreznayKapitulnik}.  They
focussed  attention on the critical value of the magnetic field, $H^\star$, at which $R_{T}(H)$ changes sign: $R_T(H)>0$ for $H<H^\star$, $R_T(H)<0$ for $H>H^\star$, and $R_T(H)\approx 0$ for $H=H^\star$.  This is often identified as the point of a SIT, despite the fact that, on both side of the ``transition,'' the $T$ dependence of the resistance is  sometimes sufficiently weak that an unbiased extrapolation to $T=0$ would yield a finite result.
Steiner {\it et al}
found that the behaviors could be sorted into two classes.  In some films, $\rho(H=H^\star) \approx h/4e^2$, and in these, a scaling collapse of the data suggestive of universal quantum critical phenomena can be achieved, and not only is $R_T(H)<0$ for $H>H^\star$, but the resistance actually grows large enough at low $T$ that it is suggestive of a truly insulating phase.  In other films,  $\rho(H=H^\star) \ll h/4e^2$, and in these,  resistance appears to approach a finite ``metallic'' value for $H$ on both sides of $H^\star$, and correspondingly any attempt to scale the data breaks down at low $T$.  Moreover, in these low resistance films, the ``critical'' resistance, $\rho(H=H^\star)$, is manifestly non-universal.  This analysis suggests that there are two possible limiting behaviors - the one in low resistance films, to which the present analysis is applicable, and that in the higher resistance films, which may, to some level of approximations, be exhibiting a superconductor to insulator transition.

However, even in the highly conducting films of Ref.~\cite{MasonKapitulnik}, experimental indications of glassy behavior has not been reported for
superconducting  films in a perpendicular magnetic field, contrary to our expectations.

\subsubsection{Transition in a parallel magnetic field in films with large $G_{2D}$}

 Most studies of superconducting films involve relatively heavy elements, such as Mo or even Pb, so that the
 spin-orbit scattering rate is substantial and $\Delta_{0}\tau_{so}>1$.  However, in Ref.  \onlinecite{Adams},
 Wu and Adams studied aluminum films where $\Delta_{0}\tau_{so}\gg 1$.  In these experiments,
  it has been observed that  in the vicinity of $H_{\|}=H^{(0)}_{c\|}$, the time dependence of the resistance
  exhibits long time relaxations with  characteristic times of order $10^{3}$ sec. During this period of time,
 the resistance changes by orders of magnitude and exhibits avalanche-like jumps. This is the characteristic
 dynamics of a glassy system. We think that this behavior is compatible with our theory, as discussed  in Section 3B.
We do not know of any experiments reported to date on the quantum transition between this superconducting glass state and the normal metal -- we believe such experiments could critically test the ideas presented here.

\subsubsection{Transition in d-wave superconductors as a function of disorder}

The cuprate high temperature superconductors are the best established example of a d-wave superconductor.
Here, the critical temperature, $T_c$, is known to vary strongly as a function of the doped hole concentration, $x$, producing two quantum critical points at which $T_c$ vanishes:  a lower critical doping concentration, $x_{1}$, on the ``underdoped'' side, and an upper critical concentration, $x_2$, on the ``overdoped'' side of the phase diagram.
  On the underdoped side of the superconducting dome,
the thermally accessible normal state above $T_c$ is
 manifestly not a good Fermi liquid.  Moreover, with increasing underdoping, these materials frequently appear to undergo a superconductor to insulator transition with a critical resistance that is typically large compared to $h/4e^2$ \cite{batlogg,ando}   Thus, the present considerations may  not be applicable.
 (However, in some instances of very high quality crystals of YBCO, the normal state revealed upon quenching superconductivity by underdoping can be somewhat metallic\cite{taillefer}.)

 It is still unclear to what extent a weak-coupling, Fermi liquid based approach is valid, even in the ``overdoped'' regime of these strongly correlated materials.  If we assume that, despite the uncertainties inherent in the strong correlation physics of the cuprates, some of the more robust of our findings apply to the cuprates as $T_c \to 0$ with overdoping, there are a number of interesting predictions we can make, none of which (to the best of our knowledge) have so far been observed experimentally.

1)  There should be a transition from a globally d-wave to a globally s-wave superconducting state at a doping concentration, $x_{d-s}$, which is less than the critical doping, $x_2$, at which $T_c$ vanishes.  While even for $x_{d-s} < x <  x_2$, any local probe will see a d-wave-like gap structure, global phase sensitive measurements should record an s-wave state.  (Some evidence of such a transition may already be present in the experiments of Ref. \onlinecite{swave}.)

2)  For $x$ near $x_2$, the superconducting state should consist of dilute puddles in which the pairing is strong, floating in an otherwise metallic sea.  (Indirect evidence of such a situation in LSCO has been presented in Ref. \onlinecite{wakimoto}.)

3)  In the metallic state with $x > x_2$, the conductivity at low temperature should diverge as $x \to x_2$, the Hall resistance should vanish, and the Wiedemann-Franz law should be increasingly strongly violated, in the sense that the conductivity should be greater than anticipated on the basis of the thermal conductivity.

Finally we would like to mention that there are various other candidates for experimental studies of the above discussed effects.   There is a growing consensus that there are multiple other examples of d-wave superconductors, including in the ``115'' family of heavy fermion superconductors and some organic superconductors.  Moreover,
Sr$_2$RuO$_4$ is known to be a p-wave superconductor. It has already been demonstrated \cite{Mackenzie} that
 superconductivity in these materials is suppressed by disorder when the parameter $k_{F}l$  of the system is still much larger than unity. Though the present theory was carried out specifically for the d-wave ({\it i.e.} spin singlet) case, we think that it is qualitatively applicable to the p-wave ({\it i.e.} spin triplet) case as well, at least in the presence of spin-orbit coupling.

We acknowledge useful discussions with M. Feigelman, D. Fisher, A. Kapitulnik, D. Khmelnitskii, B. Shklovskii, M. Skvortsov, O. Vafek, and A. P.
Young. The research was supported by NSF Grant No. DMR-0704151, and DOE Grant DE-AC02-76SF00515.

\section{Appendix: Derivation of the effective action}

We now sketch representative calculations required for the derivation of the effective actions presented in
Eqs. \ref{TDGL}, which describes the quantum dynamics of the local
superconducting ``puddles'' in the regime in which the mean-field
solution is highly inhomogeneous.
To begin with, we consider the
effective Euclidean action, $S[\Delta]$, to be a functional of  the
pair-field, $\Delta$, obtained by performing a
 Hubbard-Stratonovich transformation on an underlying microscopic
  Hamiltonian, and then integrating out the electronic degrees
of freedom. Approaching the transition from the non-superconducting
 side, we assume that the magnitude of $\Delta$ is everywhere small,
so we can expand the action in powers of $\Delta$
\ba
\label{fullaction}
S=\int d{\bf r} d {\bf r}^\prime d t d t^\prime
 \Delta^\star({\bf
r},t)K({\bf r},{\bf r}^\prime,t-t^\prime) \Delta({\bf
r}^{\prime},t^\prime) + \ldots \nonumber \ea
where
$K({\bf r},{\bf r}^\prime,t-t^\prime)$ is an appropriate imaginary time
ordered four-fermion correlation function  (which is dependent
on the precise configuration of the quenched disorder)
 and $\ldots$ represents
 higher order terms in powers of $\Delta$.
$K$ and  other response
 functions that enter the higher order terms in the effective action
  are evaluated in the normal state, {\it i.e.} they reflect
    the physics of disordered metals, not the superconducting state.

The time Fourier transform of $K$ generically has the structure,
   \begin{equation}
   \tilde K(\rr,\rrr; \omega) = K_0(\rr,\rrr) +
   |\omega| K_1(\rr,\rrr) + \ldots
   \label{Komega}
   \end{equation}
where $\ldots$ means higher order terms in powers of $\omega$. The
presence of the non-analytic $|\omega|$ dependence is generic in a
metal, and reflects the fact that in real-time, superconducting
fluctuations have an  exponential dependence on time; they decrease exponentially if
the normal metal state is stable, and increase exponentially if the normal metal is unstable.

\par In disordered systems $\tilde K(\rr,\rrr; \omega)$ is a random function
of the coordinates. Consequently near the point of the quantum phase
transition the  distribution of the order parameter can be
visualized as a sequence of superconducting puddles separated on by
a large distance.
More precisely,
  at the saddle point level, a superconducting state occurs whenever
$\hat K_0$ (by which we mean the integral operator corresponding to
$K_0$) has at least one negative eigenvalue, $\hat K_0
\Phi_{\alpha}=\ess_{\alpha} \Phi_{\alpha}$.  In other words, if
${\rm Min}[\ess_{\alpha}]$ is the smallest eigenvalue of $K_0$, then the
superconducting state   occurs when ${\rm Min}[\ess_{\alpha}]=0$ .  Generically, the smallest
eigenvalues ({\it i.e.} states deep in the ``Lifshitz tails'') are
associated with wave-functions that are spatially localized in
regions of the system that are anomalously favorable for
superconductivity.  However, the nature of these localized solutions
({\it i.e.} the spatial extent of the localized state), and the
distribution of eigenvalues in the tails of the distribution depend
on the circumstances, as we discuss in
Sections II - IV of the paper.

The full saddle-point value of $ \Delta_{sp}$, obtained by minimizing
$S[\Delta]$, can be expanded in terms of
\begin{equation}
\Delta_{sp}(\bf r) = \sum_{\alpha}
\Delta_\alpha\Phi_\alpha(\bf r) \approx  \sum_{\ess_\alpha<0}
\Delta_\alpha\Phi_\alpha(\bf r)
\label{expand}
\end{equation}

In this expansion, $\Delta_\alpha$ can be approximately interpreted
as the superconducting amplitude on puddle $\alpha$.  One trouble
with this, however, is that, like Wannier functions in a crystal,
the wave-functions $\Phi_\alpha$ are not quite as localized as they
should be, because they have small admixtures of the wave-function
from neighboring puddles which are necessitated by the orthogonality
condition, $\int d{\bf r}\Phi_\alpha^\star(\bf r)\Phi_{\alpha^\prime}(\bf r)=\delta_{\alpha,\alpha^\prime}$.

We thus obtain  Eq. \ref{TDGL} when we substitute the approximate expression in Eq. \ref{expand} into Eq. \ref{fullaction}.
The coefficient $\beta_{i}$ reflects the long-time dynamics of the
order parameter,
\begin{equation}
\beta_i=\int d{\bf r} d{\bf r}^\prime \Phi_i^\star({\bf r})
K_1({\bf r}, {\bf r}^\prime) \Phi_{i}({\bf r}^\prime).
\end{equation}
The strength of the Josephson
couplings between $i$-th and $j$-th puddles $J_{ij}$ is given by
the expression,
\begin{equation}
J_{ij}=\int d{\bf r} d{\bf r}^\prime \Phi_i^\star({\bf r})
K_0({\bf r}, {\bf r}^\prime) \Phi_{j}({\bf r}^\prime)
\label{Jij}
\end{equation}
By dimensional analysis (as well as explicit calculation) it is
clear that  $\alpha_{i}$ and $\beta_{i}$ are given by Eq.~\ref{coefficients}, while
$J_{ij}$ is given by Eq.~\ref{josephson}.


\begin{thebibliography}{99}
\bibitem{comment} 
 During the manuscript preparation, we became aware of recent
papers \cite{sachdev,voigta} which study a 
similar effective action to that in Eq. \ref{TDGL}, except without the long-range
Josephson couplings. 
These papers concluded that the critical theory is dominated by an
infnite disorder fixed 
point, which would imply universal, but unconventional critical scaling.
In the present paper, 
we have not dealt with the critical regime, so in this sense, these other
studies are
complimentary to the present one.  It will be interesting in the future
to explore the relation
between the two approaches, and in particular to explore the effects on
the critical phenomena
 of the long-range Josephson couplings and the crossover between the
small and large puddle limits, our Eqs. 1 and  4.
\end{thebibliography}
\end{document}